\documentclass[12pt]{iopart}

\usepackage[dvips]{epsfig,color}
\usepackage{amsfonts}
\usepackage{amssymb}
\usepackage{amscd}
\usepackage{pstricks}
\usepackage{amstext}
\input{xy}
\xyoption{all}
\newtheorem{example}{Example}[section]
\newtheorem{exo}[example]{Exercise}

\newtheorem{definition}[example]{Definition}
\newtheorem{proposition}[example]{Proposition}

\newtheorem{remark}[example]{Remark}
\newtheorem{no_te}[example]{Note}
\newtheorem{remarks}[example]{Remarks}

\def\sqtimes{\framebox(6.5,6.5){$\times$}\,}

\def\hs{\hbox to 3mm{}}
\def\hhs{\hbox to 5cm{}}
\def\ss{\smallskip}
\def\ms{\smallskip}
\def\bs{\bigskip}

\def\adots{\mathinner{\mkern2mu\raise1pt\hbox{.}
\mkern3mu\raise4pt\hbox{.}\mkern1mu\raise7pt\hbox{.}}}

\def\pointir{\unskip . --- \ignorespaces}

\def\up#1{\raise 1ex\hbox{\footnotesize#1}}

\def\mref#1{{\footnotesize ({\ref{#1}})}}
\def\scal#1#2{\langle #1| #2\rangle}
\def\ua{\uparrow}
\def\card{\mathrm{card}}

\def\ra{\rightarrow}

\def\ncp#1#2{#1\langle #2\rangle}

\def\LB{\mathcal{L}_{\rm Bell}}
\def\diag{\mathbf{diag}}
\def\ldiag{\mathbf{ldiag}}

\def\LDIAG{\mathbf{LDIAG}}
\def\DIAG{\mathbf{DIAG}}

\def\MQS{\mathbf{MQSym}}
\def\MON{\mathfrak{MON}}

\def\A{\mathcal{A}}
\def\B{\mathcal{B}}
\def\c{\mathcal{C}}

\def\H{{\mathcal H}}
\def\M{\mathcal{M}}
\def\St{\mathcal{S}}
\def\N{{\mathbb N}}
\def\C{{\mathbb C}}
\def\R{{\mathbb R}}
\def\Z{{\mathbb Z}}

\def\L{\mathbb{L}}
\def\V{\mathbb{V}}
\def\al{\alpha}
\def\be{\beta}

\def\ep{\epsilon}
\def\S{{\Sigma}}

\begingroup
\def\2#1{\ifnum#1<10 0\fi\the#1}
\xdef\isodayandtime{
{\2\day-\2\month-\the\year\space\2{\count0}:%
\2{\count2}}}
\endgroup

\begin{document}

\title{Hopf Algebras in General and in Combinatorial Physics: a practical introduction}

\author{G H E Duchamp$^{\dag}$, P Blasiak$^{\ddag}$, A Horzela$^\ddag$, K A Penson$^{\diamondsuit}$\\ and A I Solomon$^{\diamondsuit,\sharp}$}
\address{
$^a$ LIPN - UMR 7030\linebreak
CNRS - Universit\'e Paris 13\linebreak
F-93430 Villetaneuse, France\vspace{2mm}
}
\address
{$^\ddag$ H. Niewodnicza\'nski Institute of
Nuclear Physics, Polish Academy of Sciences\\
ul. Eliasza-Radzikowskiego 152,  PL 31342 Krak\'ow, Poland}
\address
{$^\diamondsuit$ Laboratoire de Physique Th\'eorique des Liquides,\\
Universit\'e Pierre et Marie Curie, CNRS UMR 7600\\
Tour 24 - 2e \'et., 4 pl. Jussieu, F 75252 Paris Cedex 05, France}
\address
{$^{\sharp}$ The Open University, Physics and Astronomy Department\\
Milton Keynes MK7 6AA, United Kingdom}

\eads{\linebreak\mailto{ghed@lipn-univ.paris13.fr}, \mailto{pawel.blasiak@ifj.edu.pl}, \mailto{andrzej.horzela@ifj.edu.pl}, \mailto{penson@lptl.jussieu.fr}, \mailto{a.i.solomon@open.ac.uk}\linebreak}

\begin{abstract} This  tutorial is intended to give an accessible introduction to Hopf algebras. The mathematical context is that of  representation theory, and we also illustrate the structures with examples taken from combinatorics and quantum physics, showing that in this latter case the axioms of  Hopf algebra  arise naturally. The text contains many  exercises, some  taken from physics, aimed at expanding and exemplifying the concepts introduced.
\footnote[0]{Version \today}
\end{abstract}

\section{Introduction}

Quantum Theory seen in action is an interplay of mathematical ideas and physical concepts.
From a modern  perspective its formalism and structure are founded 
on the theory of Hilbert spaces \cite{Isham,Peres}.
Using a few basic postulates, the  physical notions of a system and apparatus,
as well as transformations and measurements, are described in terms of linear operators. 
In this way the algebra of operators constitutes a proper
mathematical framework within which quantum theories may be constructed.
The structure of this algebra is determined by two operations, namely - addition and multiplication of operators; and these lie at the root of all fundamental aspects of Quantum Theory.

The formalism of quantum theory represents the physical concepts of states, observables and their transformations as objects in some Hilbert space $\mathcal{H}$ and subsequently provides a scheme for measurement predictions. Briefly, vectors in the Hilbert space describe states of a system, and linear forms in $\mathcal{V}^*$ represent basic observables. Both concepts combine in the measurement process which provides a probabilistic distribution of results and is given by the Born rule. Physical information about the system is gained by transforming the system and/or apparatus in various ways and performing measurements.
Sets of transformations usually possess some structure -- such as that of a group, semi-group, Lie algebra, etc. -- and in general can be handled within the concept of an algebra $\mathcal{A}$. The action of the algebra on the vector space of states $\mathcal{V}$ and observables $\mathcal{V}^*$ is simply its representation. Hence if an algebra is to describe physical transformations it has to have representations in all physically relevant systems. This requirement directly leads to the Hopf algebra structures in physics.

From the mathematical viewpoint the structure of the theory, modulo details, seems to be clear.
Physicists, however, need to have some additional properties and constructions to move freely in this arena. Here we will show how the structure of Hopf algebras enters into the game in the context of representations. The first issue at point is the construction of tensor product of vector spaces which is needed for the description of composite systems. 
Suppose, we know how  some transformations act on individual systems, i.e. we know representations of the algebra in each vector space $\mathcal{V}_1$ and $\mathcal{V}_2$, respectively. 
Hence natural need arises for a canonical construction of an induced representation of this algebra in $\mathcal{V}_1\otimes\mathcal{V}_2$ which would describe its action on the composite system.
Such a scheme exists and is provided by the \textit{co-product} in the algebra, i.e. a morphism $\Delta:\mathcal{A}\longrightarrow\mathcal{A}\otimes\mathcal{A}$. 
The physical plausibility of this construction requires the equivalence of representations built on  $(\mathcal{V}_1\otimes\mathcal{V}_2)\otimes\mathcal{V}_3$ and $\mathcal{V}_1\otimes(\mathcal{V}_2\otimes\mathcal{V}_3)$ -- since the composition of three systems can not depend on the order in which it is done. This requirement forces the co-product to be \textit{co-associative}. 
Another point is connected with the fact that from the physical point of view the  vector space $\mathbb{C}$ represents a trivial system having only one property -- ``being itself'' -- which can not change. Hence one should have a canonical representation of the algebra on a trivial system, denoted by $\epsilon:\mathcal{A}\longrightarrow\mathbb{C}$. Next, since the composition of any system with a trivial one can not introduce new representations, those  on $\mathcal{V}$ and $\mathcal{V}\otimes\mathbb{C}$ should be equivalent. This requirement imposes the condition on $\epsilon$ to be a \textit{co-unit} in the algebra. In this way we motivate the need for a  \textit{bi-algebra} structure in physics. The concept of an antipode enters  in the context of measurement. Measurement in a system is described in in terms of $\mathcal{V}^*\times\mathcal{V}$ and measurement predictions are given through the canonical pairing
$c:\mathcal{V}^*\times\mathcal{V}\longrightarrow \mathbb{C}$. Observables, described in the dual space $\mathcal{V}^*$,   can also be transformed and representations of appropriate algebras are given with the help of an anti-morphism $\alpha:\mathcal{A}\longrightarrow\mathcal{A}$. Physics requires that transformation preformed on the system and apparatus simultaneously should not change the measurement results, hence the pairing should trivially transform under the action of the bi-algebra. We thus obtain the condition on $\alpha$ to be an \textsl{antipode}, which is  the last ingredient of  a Hopf Algebra.

Many Hopf algebras are motivated by various theories (physical or close to physics) such as renormalization \cite{BF1,BK,B1,CK,K2}, non-commutative geometry \cite{CK,CM}, physical chemistry \cite{PC,BF2}, computer science \cite{DT}, algebraic combinatorics \cite{B1,H1,MR,CHNT}, algebra \cite{F1,F2,F3}.

\section{Operators}

\subsection{Generalities}

Throughout this text we will consider (linear) operators $\omega:V\longrightarrow V$, where
$V$ is a vector space over $k$ ($k$ is a field of scalars which can be thought of as $\mathbb{R}$ or $\mathbb{C}$). The set of all (linear) operators $V\longrightarrow V$ is an algebra (see appendix) which will be denoted by $End_\mathbb{K}(V)$.

\subsection{What is a representation}

It is not rare in Physics that we consider, instead of a single operator, a set or a family of operators $(\omega_\alpha)_{\alpha\in A}$ and often the index set itself has a structure. In the old books one finds the family-like notation, where $\rho(\alpha)$ is denoted, say, $\omega_\alpha$. As a family of operators $(\omega_\alpha)_{\alpha\in A}$ is no more than a mapping $\rho : A\mapsto End_k(V)$ (see \cite{B_Set}, Ch. II 3.4 remark), we wprefer to exhibit the mapping by considering it defined as such. This will  be precisely the concept of \textit{representation} that we will illustrate by familiar examples. Moreover using arrows allows, as we will see more clearly below, for extension and factorization procedures.  

\begin{enumerate}
\item[$\bullet$]{First case: $A$ is a group}
\end{enumerate}

\noindent
In this case, we postulate that the action of the operators be compatible with the laws of a group; that is, for all $\alpha,\beta\in A$,

\begin{eqnarray}\label{group_rep1}
\left\{\begin{array}{l}
\rho(\alpha.\beta)=\rho(\alpha)\circ\rho(\beta)\\
\rho(\alpha^{-1})=(\rho(\alpha))^{-1} 
\end{array}\right.
\end{eqnarray}

which is equivalent to 

\begin{eqnarray}\label{group_rep2}
\left\{\begin{array}{l}
\rho(\alpha.\beta)=\rho(\alpha)\circ\rho(\beta)\\
\rho(1_A)=1_{End(V)}
\end{array}\right.
\end{eqnarray}

Note that each of these conditions implies that the range of $\rho$ is in $Aut_k(V)$ ($=GL_n(k)$, the linear group), the set of one-to-one elements of $End_k(V)$ (called automorphisms). 

\begin{enumerate}
\item[$\bullet$]{Second case: $A$ is a Lie algebra}
\end{enumerate}

\noindent
In this case, one requires that
\begin{eqnarray}\label{Lie_rep}
\rho([\alpha,\beta])=\rho(\alpha)\circ\rho(\beta)-\rho(\beta)\circ\rho(\alpha)=[\rho(\alpha),\rho(\beta)].
\end{eqnarray}

\bs
We will see that these two types of action (of a group or Lie algebra) can be unified through the concept of the representation of an algebra (or which amounts to the same thing, of a module). 

\ss
In the first case, one invokes the \emph{group algebra} $k[A]$ (see appendix). In the case of a Lie algebra, one invokes the \emph{enveloping algebra} $\mathcal{U}(A)$ (or $\mathcal{U}_k(A)$ see appendix). 
In both cases, the original representation $\rho$ is extended to a representation of an \emph{associative algebra with unit} (AAU) as follows: 

\begin{eqnarray}\nonumber
\ \ \ \ \ \ \xymatrix{
G\ar[r]^{\rho}\ar[d]^{can}&End_k(V)\\
 k[G]\ar[ur]_{\hat{\rho}} &\\
}
&&\ \ \ \ \ \ \ \ \ \ \ \ \ \ \ \ \ \ \ \ \ \ 
\xymatrix{
\mathfrak{G}\ar[r]^{\rho}\ar[d]^{can}&End_k(V)\\
 \mathcal{U}_k(\mathfrak{G})\ar[ur]_{\hat{\rho}} &\\
}\end{eqnarray}
So far we have not defined what a representation of AAU is.
Keeping the philosophy of \mref{group_rep1} (or \mref{group_rep2}) and \mref{Lie_rep}, we can state the following definition:

\begin{definition} Let $(\A,+,\cdot)$ be an AAU. A collection of operators $\{\rho(\alpha)\}_{\alpha\in \A}$ in a vector space $V$ is said to be a representation of $\A$ iff the mapping $\rho : \A \mapsto End(V)$ is compatible with the operations and units of $A$. This means that, identically (i. e. for\ all $\alpha,\beta\in \A$ and $\lambda\in\mathbb{K}$).
\begin{eqnarray}
\left\{\begin{array}{l}
\rho(\alpha+\beta)=\rho(\alpha)+\rho(\beta),\ \ \ \ \ \ 
\rho(\lambda\alpha)=\lambda\rho(\alpha),\\
\rho(\alpha\cdot\beta)=\rho(\alpha)\circ\rho(\beta),\\
\rho(1_\A)=Id_V,
\end{array}\right.
\end{eqnarray}
where $\circ$ denotes composition of operators.
\end{definition}

\begin{remarks}
(i) This is equivalent to saying that the arrow $\rho : \A \mapsto End(V)$
from $\A$ to $End(V)$ is a morphism of algebras (with units).

(ii) In this case, it is sometimes convenient to denote by $\alpha.v$ the action of $\rho(\alpha)$ on $v$ (i.e. the element $\rho(\alpha)[v]$) for $\alpha\in \A$ and $v\in V$.

(iii) It may happen (and this often occurs) that a representation has relations that are not present in the original algebra. In this case the representation is said to be not \emph{faithful}. More rigourously a representation is said to be faithful iff $\rho$ is injective or, which is equivalent, $ker(\rho)=\rho^{-1}(\{0\})=\{0\}$ (for algebras and Lie algebras) and $ker(\rho)=\rho^{-1}(\{1_V\})=\{1_G\}$ (for groups).
\end{remarks}

\begin{example} : Let $G=\{1,c,c^2\}$ be the cyclic group of order 3 ($c$ is the cycle $1\rightarrow 2\rightarrow 3\rightarrow 1$), $G$ admits the plane representation by 
\begin{eqnarray}
\hspace{-2.4cm}\rho(c)=\left(\begin{array}{cc}-1/2&-\sqrt{3}/2\\\sqrt{3}/2&-1/2\end{array}\right)\ \text{(it is the matrix corresponding to a rotation of $2\pi/3$)}.
\end{eqnarray}
Thus,
\begin{eqnarray}\rho(c^2)=\left(\begin{array}{cc}-1/2&\sqrt{3}/2\\-\sqrt{3}/2&-1/2\end{array}\right) \ \ \ \ \text{and, of course,}\ \ \ \ \rho(1)=\left(\begin{array}{cc}1&0\\0&1\end{array}\right). 
\end{eqnarray}
The representation $\rho$ is faithful while its extension to the group algebra is not, as seen from:
\begin{eqnarray}
\rho(1+c+c^2)=\left(\begin{array}{cc}0&0\\0&0\end{array}\right) \text{whereas}\ 1+c+c^2\neq0\ \text{in}\ \mathbb{C}[G].
\end{eqnarray}
\end{example}

\begin{no_te}
Note that the situation is even worse for a Lie algebra, as $\mathcal{U}_k(\mathfrak{G})$ is infinite dimensional iff $\mathfrak{G}$ is not zero.
\end{no_te}

\section{Operations on representations}
Now, we would like to see the representations of AAU as building blocks to construct new ones. The elementary operations on vector spaces are:

\begin{center}
\begin{itemize}
\item[-] sums 
\item[-] tensor products
\item[-] duals
\end{itemize}
\end{center}

Hence, an important problem is:\\ 
Given representations $\rho_i:\A\mapsto V_i$ on the building blocks $V_i\ ;\ i=1,2$ how does one  naturally construct representations on $V_1\oplus V_2$, $V_1\otimes V_2$ and $V_i^*$. 

Sums will cause no problem as the sum $V_1\oplus V_2$ of two vector spaces $V_1$ and $V_2$ amounts to taking their cartesian product $V_1\oplus V_2\cong V_1\times V_2$. Then, if 
$\rho_i : \A\mapsto V_i\ ;\ i=1,2$ are two representations of $\A$ then the mapping $\rho_1\oplus\rho_2:\A\mapsto V_1\otimes V_2$ such that
\begin{equation}
\rho_1\oplus\rho_2(a)[(v_1,v_2)]=(\rho_1(a)[v_1],\rho_2(a)[v_2])
\end{equation}
which can be symbolically written
\begin{equation}
\rho_1\oplus\rho_2=
\left(
{\begin{array}{cc}
\rho_1 & 0 \\
0 & \rho_2
\end{array}}
 \right),
\end{equation}
is a representation of $\A$ in $V_1\oplus V_2$.

\ss
Dualization will be discussed later and solved by the existence of an \textit{antipode}. Now, we start with the problem of constructing representations on tensor products. This will be solved by means of the notion of ``scheme of actions'' which is to be formalized, in our case, by the concept of comultiplication (or coproduct).

\subsection{Arrows and addition or multiplication formulas}

Let us give first some examples where  comultiplication naturally arises.\\
We begin with functions admitting an ``addition formula'' or ``multiplication formula''. This means functions such that for all $x$, $y$
\begin{eqnarray}\label{addition_formula}
f(x\ast y)=\sum_{i=1}^n f_i^{(1)}(x)f_i^{(2)}(y),
\end{eqnarray}
where $\ast$ is a certain (associative) operation on the defining set of $f$ and $(f_i^{(1)},f_i^{(2)})_{i=1}^n$ be two (finite) families of functions on the same set (see exercises \mref{complex_rep_func} and \mref{rep_func} on representative functions).\\ 
The first examples are taken in the function space\footnote{As usual, $Y^X$ is the set of all mappings from $X$ to $Y$, see appendix.} $\R^\R$ (with $*=+$, the ordinary addition of real numbers). The following functions admit ``addition formulas'' which can be expressed diagramatically as follows. 

\bigskip
\begin{center}
\begin{tabular}{|c|c|}
\hline
Diagram & Addition formula\\
\hline\hline
$\xymatrix{
\ar[r]^{\ \ +}\mathbb{R}\times\mathbb{R}\ar[dr]_{\cos_1\cos_2-\sin_1\sin_2\ }&\ar[d]^{\cos}\mathbb{R}\\
 &\mathbb{R}\\
}$
&$\cos(x+y)=\cos(x)\cos(y)-\sin(x)\sin(y)$\\
\hline
$\xymatrix{
\ar[r]^{\ \ +}\mathbb{R}\times\mathbb{R}\ar[dr]_{\sin_1\cos_2+\sin_2\cos_1\ }&\ar[d]^{\sin}\mathbb{R}\\
 &\mathbb{R}\\
}$
&$\sin(x+y)=\sin(x)\cos(y)+\sin(y)\cos(x)$\\
\hline
$\xymatrix{
\ar[r]^{\ \ +}\mathbb{R}\times\mathbb{R}\ar[dr]_{\exp_1\exp_2 }&\ar[d]^{\exp}\mathbb{R}\\
 &\mathbb{R}\\
}$
&$\exp(x+y)=\exp(y)\exp(x)$\\
\hline
$\xymatrix{
\ar[r]^{\ \ +}\mathbb{R}\times\mathbb{R}
\ar[dr]_{\sum_{j=0}^n {n \choose j} pr_1^{(j)}\ pr_2^{(n-j)}\ \ } & \ar[d]^{(\ )^n}\mathbb{R}\\
 &\mathbb{R}\\
}$
&$(x+y)^n=\sum_{j=0}^n {n \choose j} x^j\ y^{n-j}$\\
\hline
\end{tabular}
\end{center}

\bigskip

Another example can be given where the domain set (source) is $\mathbb{C}^{n\times n}$, the algebra of square $n\times n$ matrices with complex coefficients. Let $a_{ij}:\mathbb{C}^{n\times n}\longrightarrow\mathbb{C}$ be the linear form which ``takes'' the coefficient of address $(i,j)$ (row $i$ and column $j$), that is to say $a_{ij}(M):=M[i,j]$. Then,  the law of multiplication of matrices says that $MN[i,j]=\sum_{k=1}^n M[i,k]N[k,j]$, which can be represented in the style of Eq. \mref{addition_formula} by

\begin{eqnarray}\label{matrixcoprod}
a_{ij}(MN)=\sum_{k=1}^n a_{ik}(M)a_{kj}(N).
\end{eqnarray}

\begin{eqnarray}\nonumber
\begin{array}{c}\xymatrix{
\ar[rr]^{\ \ \ \ product}\mathbb{C}^{n\times n}\times\mathbb{C}^{n\times n}\ar[drr]_{\sum_{k=1}^n (a_{ik})_1(a_{kj})_2\ \  \ \ \ \ }&&\ar[d]^{a_{ij}}\mathbb{C}^{n\times n}\\
 &&\mathbb{C}\\
}
\end{array}
\end{eqnarray}

\begin{remark} Note that formula \mref{matrixcoprod} holds when the definition set (source) is a (multiplicative) semigroup of matrices (for example, the semigroup of unipotent positive matrices). 
\end{remark}

We  now proceed to linear mappings that admit such ``addition'' or, rather, ``multiplication'' formulas.

\bs\noindent
\textbf{\emph{Derivations}}: Let $\mathcal{A}$ be an arbitrary algebra with law of multiplication:
\begin{eqnarray}
\mathcal{A}\otimes\mathcal{A}\stackrel{\mu}{\longrightarrow}\mathcal{A}.
\end{eqnarray} 
A derivation of $\mathcal{A}$ is an operator $D:\mathcal{A}\longrightarrow\mathcal{A}$ which follows the Leibniz rule, that is for all $x,y\in\mathcal{A}$, one has
\begin{eqnarray}
D(xy)=D(x)y+xD(y)\ \ \ \ \text{(Leibniz rule)}.
\end{eqnarray}
In the spirit of what has been represented above one has
\begin{eqnarray}\nonumber
\xymatrix{
\ar[r]^{\ \ \mu}\mathcal{A}\otimes\mathcal{A}\ar[dr]_{D_1Id_2+Id_1D_2}&\ar[d]^{D}\mathcal{A}\\
 &\mathcal{A}
}
\end{eqnarray}
which (as we have linear spaces and mappings) can be better represented by 
\begin{eqnarray}\nonumber
\xymatrix{
\ar[r]^{\ \ \mu}\mathcal{A}\otimes\mathcal{A}\ar[d]_{D\otimes Id+Id\otimes D}&\ar[d]^{D}\mathcal{A}\\
 \mathcal{A}\otimes\mathcal{A}\ar[r]^{\ \ \mu}&\mathcal{A}
}
\end{eqnarray}

\bs\noindent
\textbf{\emph{Automorphisms}}:
An automorphism of $\mathcal{A}$ is an invertible linear mapping  $g:\mathcal{A}\longrightarrow\mathcal{A}$ such that for all $x,y\in\mathcal{A}$, one has
\begin{eqnarray}
g(xy)=g(x)g(y),
\end{eqnarray}
which, in the spirit of what precedes can be represented by 
\begin{eqnarray}\nonumber
\xymatrix{
\ar[r]^{\ \ \mu}\mathcal{A}\otimes\mathcal{A}\ar[d]_{g\otimes g}&\ar[d]^{g}\mathcal{A}\\
 \mathcal{A}\otimes\mathcal{A}\ar[r]^{\ \ \mu}&\mathcal{A}
}
\end{eqnarray}
Now remark that, classically, group representations act as automorphisms and representations of Lie algebras act as derivations. This immediately provides  a scheme for constructing tensor products of two representations.

\bs\noindent
\textbf{\emph{Tensor product of two representations (groups and Lie algebras)}}:
First, take two representations of a group $G$, $\rho_i:G\longrightarrow End(V_i)$, $i=1,2$. The action of $g\in G$ on the tensor space $V_1\otimes V_2$ is given by
\begin{eqnarray}
g(v_1\otimes v_2)=g(v_1)\otimes g(v_2).
\end{eqnarray}
This means that the ``tensor product'' of the two (group) representations $\rho_i,\ i=1,2$ is given by the following data
\begin{itemize}
	\item Space : $V_1\otimes V_2$
	\item Action : $\rho_1\sqtimes \rho_2 : g\ra \rho_1(g)\otimes \rho_2(g)$
\end{itemize}
  
Likewise, if we have two representations $\rho_i:\mathfrak{G}\longrightarrow End(V_i)$, $i=1,2$ of the Lie algebra $\mathfrak{G}$ the action of $g\in \mathfrak{G}$ on a tensor product $V_1\otimes V_2$ is given by
\begin{eqnarray}
g(v_1\otimes v_2)=g(v_1)\otimes v_2+v_1\otimes g(v_2).
\end{eqnarray}
Again, the ``tensor product'' of the two (Lie algebra) representations $\rho_i,\ i=1,2$ is given by the following data
\begin{itemize}
	\item Space : $V_1\otimes V_2$
	\item Action : $\rho_1\sqtimes \rho_2 : g\ra \rho_1(g)\otimes Id_{V_2}(g)+Id_{V_1}(g)\otimes \rho_2(g)$
\end{itemize}

Roughly speaking, in the first case $g$ acts by $g\otimes g$ and in the second one by $g\otimes 1+1\otimes g$. 

\noindent In view of two above cases it is convenient to construct linear mappings:
\begin{eqnarray}
\mathcal{A}\stackrel{\Delta}{\longrightarrow}\mathcal{A}\otimes\mathcal{A}\ ,
\end{eqnarray} 
such that, in each case, $\rho_1\sqtimes\rho_2=(\rho_1\otimes\rho_2)\circ\Delta$.

\noindent In the first case ($\mathcal{A}=\mathbb{C}[G]$) one gets
\begin{eqnarray}
\Delta\left(\sum_{g\in G}\alpha_g g\right)=\sum_{g\in G}\alpha_g g\otimes g.
\end{eqnarray} 
In the second case, one has first to construct the comultiplication on the monomials\\ 
$g_1...g_n; g_i\in \mathfrak{G})$ as they span ($\mathcal{A}=\mathcal{U}_k(\mathfrak{G})$). Then, using the rule $\Delta(g)=g\otimes 1+1\otimes g$ (for $g\in \mathfrak{G}$) and the fact that $\Delta$ is supposed to be a morphism for the multiplication (the justification of this rests on the fact that the constructed action must be a representation see below around formula \mref{Delta_morph} and exercise \mref{tensor_rep}), one has 
\begin{eqnarray}
\Delta(g_1...g_n)&=&(g_1\otimes 1+1\otimes g_1)(g_2\otimes 1+1\otimes g_2)...(g_n\otimes 1+1\otimes g_n)\cr
&=&\sum_{I+J=[1...n]}g[I]\otimes g[J].
\end{eqnarray} 
Where, for $I=\{i_1,i_2,\cdots ,i_k\}$ ($1\leq i_1<i_2<\cdots <i_k\leq n$), $g[I]$ stands for 
$g_{i_1}g_{i_2}\cdots g_{i_k}$
In each case (group algebra and envelopping algebra) one again gets a mapping $\Delta : \A\mapsto \A\otimes \A$ which will be expressed by 
\begin{eqnarray}
\Delta(a)=\sum_{i=1}^n a_i^{(1)}\otimes a_i^{(2)}
\end{eqnarray}
which is rephrased compactly by
\begin{eqnarray}\label{sweedlers_notation}
\Delta(a)=\sum_{(1)(2)} a_{(1)}\otimes a_{(2)}.
\end{eqnarray}

The action of $a\in \A$ on a tensor $v_1\otimes v_2$ is then, in both cases, given by
\begin{eqnarray}\label{tensor_act}
a.(v_1\otimes v_2)=\sum_{i=1}^n a_i^{(1)}. v_1\otimes a_i^{(2)}. v_2=\sum_{(1)(2)} a_{(1)}. v_1\otimes a_{(2)}. v_2.
\end{eqnarray}

One can easily check, in these two cases, that 
\begin{equation}\label{tensor_act1}
a.b.(v_1\otimes v_2)=(ab).(v_1\otimes v_2),	
\end{equation}
but in general \mref{tensor_act} does not guarantee \mref{tensor_act1}; this point will be discussed below in section \mref{reasonable_tensor}.    

\noindent Expression \mref{sweedlers_notation} is very convenient for proofs and computations and known as Sweedler's notation.

\begin{remarks} i) In every case, we have extracted the ``scheme of action'' for building the tensor product of two representations. This scheme (a linear mapping $\Delta : \A\mapsto \A\otimes \A$) is independent of the considered representations and, in each case,  
\begin{equation}\label{tensor1}
	\rho_1\sqtimes \rho_2=(\rho_1\otimes \rho_2)\circ \Delta
\end{equation}
ii) Sweedler's notation becomes transparent when one speaks the language of ``structure constants''. Let $\Delta: \c\mapsto \c\otimes \c$ be a comultiplication and $(b_i)_{i\in I}$ a (linear) basis of $\c$. One has
\begin{equation}\label{struct_const}
\Delta(b_i)=\sum_{j,k\in I}\lambda_i^{j,k}\ b_j\otimes b_k\ .	
\end{equation}
the family $(\lambda_i^{j,k})_{i,j,k\in I}$ is called the ``structure constants'' of the comultiplication $\Delta$. Note the duality with the notion the ``structure constants'' of a multiplication\\ 
$\mu : \A\otimes \A \mapsto \A$ : if $(b_i)_{i\in I}$ is a (linear) basis of $\A$, one has 
\begin{equation}\label{struct_const}
\mu(b_i\otimes b_j)=\sum_{k\in I}\lambda_{i,j}^k\ b_k\ .	
\end{equation}
For necessary and sufficient conditions for a family to be structure constants (see exercise \mref{struct_const_ex}).
\end{remarks}

\bs
Then, the general construction for tensor products goes as follows.

\begin{definition}: Let $\mathcal{A}$ be a vector space, a comultiplication $\Delta$ on $\mathcal{A}$ is a linear mapping 
$$
\mathcal{A}\stackrel{\Delta}{\longrightarrow}\mathcal{A}\otimes\mathcal{A}.
$$

Such a pair (vector space, comultiplication) without any prescription about the linear mapping ``comultiplication'' is called a coalgebra.
\end{definition}

Now, imitating \mref{tensor1}, if $\mathcal{A}$ is an algebra and $\rho_1$, $\rho_2$ are representations of $\mathcal{A}$ in $V_1$, $V_2$, for each $a\in \A$, we can construct an action of $a$ on $V_1\otimes V_2$ by
\begin{eqnarray}
V_1\otimes V_2\stackrel{(\rho_1\otimes \rho_2)\circ \Delta(a)}{\longrightarrow}V_1\otimes V_2.
\end{eqnarray}
This means that if $\Delta(a)=\sum_{(1)(2)} a_{(1)}\otimes a_{(2)}$, then, ($a$) acts on the tensor product by  
\begin{eqnarray}\label{tensor_action}
\hspace{-1cm} a.(v_1\otimes v_2)=\sum_{(1)(2)} a_{(1)}. v_1\otimes a_{(2)}. v_2=\sum_{(1)(2)} \rho_1(a_{(1)})[v_1]\otimes \rho_2(a_{(2)})[v_2].
\end{eqnarray}

\noindent But, at this stage, it is just an action and not (necessarily) a representation of $\mathcal{A}$. We shall later give the requirements on $\Delta$ for the construction of the tensor product to be reasonable (i. e. compatible with the usual tensor properties). 

For the moment let us pause and consider some well known examples of comultiplication.

\subsection{Combinatorics of some comultiplications}

The first type of comultiplication is given by duality. This means by a formula of type 
\begin{eqnarray}\label{dual_law}
\scal{\Delta(x)}{y\otimes z}^{\otimes2}=\scal{x}{y\ast z}
\end{eqnarray}
for a certain law of algebra $V\otimes V\stackrel{\ast}{\mapsto} V$, where $\scal{\ }{\ }$ is a non degenerate scalar product in $V$ and $\scal{\ }{\ }^{\otimes2}$ stands for its extension to $V\otimes V$. In the case of words $\ast$ is the concatenation and $\scal{\ }{\ }$ is given by 
$\scal{u}{v}=\delta_{u,v}$. The comultiplication $\Delta_{Cauchy}$, dual to the concatenation, is given on a word $w$ by
\begin{eqnarray}
\Delta(w)=\sum_{uv=w}u\otimes v.
\end{eqnarray}
In the same spirit, one can define a comultiplication on the algebra of a finite group by 

\begin{equation}
\Delta(g)=\sum_{g_1g_2=g}g_1\otimes g_2.	
\end{equation}

The second example is given by the multiplication law of elementary comultiplications, that is, if for each letter $x$ one has
$\Delta(x)=x\otimes 1+1\otimes x$, then
\begin{eqnarray}
\Delta(w)&=&\Delta(a_1...a_n)=\Delta(a_1)\Delta(a_2)...\Delta(a_n)\cr
 &=&\sum_{I+J=[1...n]}w[I]\otimes w[J]
\end{eqnarray}
where $w[\{i_1,i_2,\cdots i_k\}]=a_{i_1}a_{i_2}\cdots a_{i_k}$ (for $1\leq i_1<i_2<\cdots <i_k\leq n$). This comultiplication is dual to \mref{infiltr} below for $q=0$ (shuffle product).\\ 
Another example is a deformation (perturbation for small $q$) of the preceding. With $\Delta(a)=a\otimes 1+1\otimes a+ q a\otimes a$, one has
\begin{eqnarray}
\Delta(w)&=&\Delta(a_1...a_n)=\Delta(a_1)\Delta(a_2)...\Delta(a_n)\cr
 &=&
\sum_{I\cup J=[1...n]} q^{|I\cap J|} w[I]\otimes w[J].
\end{eqnarray}

Note that this comultiplication is dual (in the sense of \mref{dual_law}) to the $q$-infiltration product given by the recursive formula (for general $q$ and with $1_{A^*}$ as the empty word)

\begin{eqnarray}\label{infiltr}
\hspace{-15mm} w\ua 1_{A^*}=1_{A^*}\ua w=w\cr 
\hspace{-15mm} au\ua bv=a(u\ua bv)+ b(au\ua v)+ q\delta_{a,b} (u\ua v).
\end{eqnarray}

This product is an interpolation between the shuffle ($q=0$) and the (classical) infiltration ($q=1$) \cite{DFLL}.

\section{Requirements for a reasonable construction of tensor products}\label{reasonable_tensor}

We have so far constructed an action of $\A$ on tensors, but nothing indicates that this is a representation (see exercise \mref{tensor_rep}). So, the following question is natural.

\bs
{\bf Q.1.)} If $\A$ is an algebra and $\Delta :\A\mapsto \A\otimes\A$, what do we require on $\Delta$ if we want the construction above to be a representation of $\A$ on tensor products ?

\bs
For $a,b\in \A$, $\rho_i$ representations of $\mathcal{A}$ in $V_i$, and $v_i\in V_i$ for $i=1,2$, we must have the following identity:
\begin{eqnarray}\label{action}
\!\!\!\!\!\!\!\!\!\! a.(b.v_1\otimes v_2)=(ab).v_1\otimes v_2\ \Longleftrightarrow\ \Delta(ab).v_1\otimes v_2=\Delta(a).(\Delta(b).v_1\otimes v_2).
\end{eqnarray}
One can prove that, if this is true identically for all $a,b\in \mathcal{A}$ and all pairs of representations (see exercise \mref{tensor_rep}), one has 
\begin{eqnarray}\label{Delta_morph}
\Delta(ab)=\Delta(a)\Delta(b).
\end{eqnarray}
and, of course, if the latter holds, \mref{action} is true.\\
This can be rephrased by saying that $\Delta$ is a morphism $\A\mapsto \A\otimes \A$.

\ss 
Now, one would like to keep compatibility with the associativity of tensor products. This means that if we want to tensor $u\otimes v$ with $w$ it must give the same action as tensoring $u$ with $v\otimes w$. This means that we have to address the following question.\\

\bs
{\bf Q.2.)} If $\A$ is an algebra and $\Delta :\A\mapsto \A\otimes\A$ is a morphism of algebras, what do we require on $\Delta$ if we want the construction above to be associative ? 

\bs
More precisely, for three representations $\rho_i,\ i=1,2,3$ of $\A$, we want 
\begin{equation}\label{tensor_rep_assoc}
\rho_1\sqtimes(\rho_2\sqtimes\rho_3)=(\rho_1\sqtimes\rho_2)\sqtimes\rho_3	
\end{equation}
up to the identifications $(u\otimes v)\otimes w=u\otimes (v\otimes w)=u\otimes v\otimes w$ (if one is not satisfied with this identification, see exercise \mref{tensor_assoc}). 

Let us compute (up to the identification above)
\begin{eqnarray}
\hspace{-1cm} a.[(u\otimes v)\otimes w]=\Delta(a).(u\otimes v)\otimes w=\Big((\Delta\otimes Id)\circ \Delta(a)\Big)\Big(u\otimes v\otimes w\Big)
\end{eqnarray}
on the other hand
\begin{eqnarray}
\hspace{-1cm} a.[u\otimes (v\otimes w)]=\Delta(a).u\otimes (v\otimes w)=\Big((Id\otimes \Delta)\circ\Delta(a)\Big)\Big(u\otimes v\otimes w\Big).
\end{eqnarray}
Again, one can prove (see exercise \mref{tensor_assoc}) that this holds identically (i. e. for every $a\in \A$ and triple of representations) iff $(Id\otimes \Delta)\circ\Delta=(\Delta\otimes Id)\circ\Delta$, \emph{i.e.}
\begin{eqnarray}\label{co-associativity}
\xymatrix{
\ar[rr]^{\Delta}\mathcal{A}\ar[d]_{\Delta}&&\ar[d]^{Id\otimes \Delta}\mathcal{A}\otimes\mathcal{A}\\
 \mathcal{A}\otimes\mathcal{A}\ar[rr]^{\Delta\otimes Id}&&\mathcal{A}\otimes\mathcal{A}\otimes\mathcal{A}
}
\end{eqnarray}

\bs
\begin{remark} 
The property \mref{co-associativity} is called co-associativity since if one reverses the arrows and replaces $\Delta$ by $\mu$, the multiplication in an algebra, the diagram expresses associativity (see also exercise \mref{coalgebra_dual} on duals of co-algebras).
\end{remark}

\begin{eqnarray}\nonumber
\xymatrix{
\ar[rr]^{\mu\otimes Id}\mathcal{A}\otimes\mathcal{A}\otimes\mathcal{A}\ar[d]_{Id\otimes\mu}&&\ar[d]^{\mu}\mathcal{A}\otimes\mathcal{A}\\
 \mathcal{A}\otimes\mathcal{A}\ar[rr]^{\mu}&&\mathcal{A}
}
\end{eqnarray}

\vspace{1cm}
But the tensor product is not only associative, it has a ``neutral'' map which is ``tensoring by the field of scalars''. This derives from the fact that the canonical mappings 
\begin{eqnarray}\label{can_rl}
V\otimes_k k\stackrel{can_r}{\longrightarrow}V\stackrel{can_l}{\longleftarrow}k\otimes_k V.
\end{eqnarray}
This can be summarized by the following question.

\bs
{\bf Q.3.)} If $\A$ is an algebra and $\Delta :\A\mapsto \A\otimes\A$ a co-associative morphism of algebras, what do we require on $\Delta$ if we want the construction above to admit ``tensoring by the field of scalars'' as neutral ? 

\bs
More precisely, we must have a representation of $\mathcal{A}$ in $\mathbb{C}$ (which means a morphism of algebras $\mathcal{A}\stackrel{\epsilon}{\longrightarrow}\mathbb{C}$) such that for a representation $\rho$ of $\A$, we want 
\begin{equation}
\rho\sqtimes \ep=\ep\sqtimes \rho=\rho	
\end{equation}
up to the identification $u\otimes 1=1\otimes u=u$ through the isomorphisms \mref{can_rl} (if one is not satisfied with this identification, see exercise \mref{tensor_unit}).\\ 
Hence, for all $a\in\mathcal{A}$ and $\rho$ representation on $V$, we should have:
\begin{eqnarray}
can_r( a.(v\otimes 1))=a.v,
\end{eqnarray}
and
\begin{eqnarray}
a.(v\otimes 1)&=&\Big(\sum_{(1)(2)}\rho(a_{(1)})\otimes\epsilon(a_{(2)})\Big)[v\otimes 1]\nonumber\\
&=&(\rho\otimes Id_\C)\circ\Big(\sum_{(1)(2)}a_{(1)}\otimes\epsilon(a_{(2)})\Big)[v\otimes 1]\\\nonumber
&=&(\rho\otimes Id_\C)\circ(Id\otimes\epsilon)\circ\Delta(a)[v\otimes 1],\\\nonumber&&\\
\ \ \ \ a.v&=&\rho(a)[v]=can_r(\rho(a)[v]\otimes1)=can_r(\rho\otimes Id(a)[v\otimes1]).
\end{eqnarray}
Similar computations could be made on the left, we leave them to the reader as an exercise.

\ss
This means that one should require that 
\begin{eqnarray}\nonumber
\xymatrix{
\ar[rr]^{\Delta}\mathcal{A}\ar[d]_{Id}&&\ar[d]^{Id\otimes\epsilon}\mathcal{A}\otimes\mathcal{A}\\
\mathcal{A}&&\ar[ll]_{can_r}\mathcal{A}\otimes\mathbb{C}
}
&\ \ \ \ \ \ \ \ \ \ \ &\xymatrix{
\ar[rr]^{\Delta}\mathcal{A}\ar[d]_{Id}&&\ar[d]^{\epsilon\otimes Id}\mathcal{A}\otimes\mathcal{A}\\
\mathcal{A}&&\ar[ll]_{can_l}\mathbb{C}\otimes\mathcal{A}
}
\end{eqnarray}
Such a mapping $\epsilon:\mathcal{A}\longrightarrow\mathbb{C}$ is called a {\em co-unit}.

\begin{remark} Again, one can prove (see excercice \mref{coalgebra_dual} for details) that
\begin{eqnarray}
\epsilon\text{ is a counit for }(\A,\Delta)\ \ \Longleftrightarrow\ \  \epsilon\text{ is a counit for } (\A^*,*_\Delta).
\end{eqnarray}
\end{remark}

\section{Bialgebras}

Motivated by the preceding discussion, we define a {\em bialgebra} an algebra (associative with unit) endowed with a comultiplication (co-associative with counit) which allows for the two tensor properties of associativity and unit (see discussion above). More precisely  

\begin{definition}: $(\mathcal{A},\cdot,1_\mathcal{A},\Delta,\epsilon)$ is said to be a bialgebra iff
\begin{enumerate}
\item[(1)]{$(\mathcal{A},\cdot,1_\mathcal{A})$ is an AAU,}

\item[(2)]{$(\mathcal{A},\Delta,\epsilon)$ is a coalgebra coassociative with counit,}

\item[(3)]{$\Delta$ is a morphism of AAU and $\epsilon$ is a morphism of AAU.}
\end{enumerate}
\end{definition}

The name bialgebra comes from the fact that the space $\A$ is endowed with two structures (one of AAU and one of co-AAU) with a certain compatibility between the two.

\subsection{Examples of bialgebras}

\subsection*{Free algebra (word version : noncommutative polynomials)}

Let $A$ be an alphabet (a set of variables) and $A^*$ be the free monoid constructed on $A$ (see Basic Structures \mref{basic_structures}). For any field of scalars $k$ (one may think  of $k=\R$ or $\C$), we call the algebra of noncommutative polynomials $\ncp{k}{A}$ (or free algebra), the algebra $k[A^*]$ of the free monoid $A^*$ constructed on $A$. This is the set of functions $f : A^*\mapsto k$ with finite support endowed with the convolution product 
\begin{equation}
	f\ast g(w)=\sum_{uv=w}f(u)g(v)
\end{equation}

Each word $w\in A^*$ is identified with its characteristic function (i.e. the Dirac function with value $1$ at $w$ and $0$ elsewhere). These functions form a basis of $\ncp{k}{A}$ and then, every $f\in \ncp{k}{A}$ can be written uniquely as a finite sum $f=\sum f(w)w$.\\
The inclusion mapping $A\hookrightarrow \ncp{k}{A}$ will be denoted here by $can_A$.  

\textbf{Comultiplications} The free algebra $\ncp{k}{A}$ admits many comultiplications (even with the two requirements of being a morphism and coassociative). As $A^*$ is a basis of $\ncp{k}{A}$, it is sufficient to define it on the words (if we require $\Delta$ to be a morphism it is enough to define it on letters).\\
\textit{Example 1}\pointir The first example is the dual of the Cauchy (or convolution) product
\begin{equation}
	\Delta(w)=\sum_{uv=w} u\otimes v
\end{equation}
is not a morphism as 
$$
\Delta(ab)=ab\otimes 1+a\otimes b+ 1\otimes ab
$$ 
and 
$$
\Delta(a)\Delta(b)=ab\otimes 1+a\otimes b+ b\otimes a+1\otimes ab
$$  
but it can be checked that it is coassociative (see also exercice \mref{coalgebra_dual} for a quick proof of this fact).

\ss
\textit{Example 2}\pointir A second example is given, on the alphabet $A=\{a,b\}$ by 
$$
\Delta(a)=a\otimes b;\ \Delta(b)=b\otimes a
$$ 
then $\Delta(w)=w\otimes \bar{w}$ where $\bar{w}$ stands for the word $w$ with $a$ (resp. $b$) changed in $b$ (resp. $a$). This comultiplication is a morphism but not coassociative as 
$$
(I\otimes \Delta)\circ\Delta(a)=a\otimes b\otimes a\ ;\ (\Delta\otimes I)\circ\Delta(a)=a\otimes a\otimes b
$$  

\textit{Example 3}\pointir The third example is given on the letters by 
$$
\Delta(a)=a\otimes 1+1\otimes a+q a\otimes a
$$ 
where $q\in k$. One can prove that
\begin{equation}
\hspace{-1cm}	\Delta(w)=\Delta(a_1...a_n)=\Delta(a_1)\Delta(a_2)...\Delta(a_n)=
\sum_{I\cup J=[1..|w|]} q^{|I\cap J|}w[I]\otimes w[J]
\end{equation}
this comultiplication is coassociative.

For $q=0$, one gets a comultiplication given on the letters by $\Delta_s(a)=a\otimes 1+1\otimes a$. For every polynomial $P\in \ncp{k}{A}$, set $\ep(P)=P(1_{A^*})$ (the constant term). Then $(\ncp{k}{A},\ast,\Delta_s,\ep)$ is a bialgebra. 

One has also another bialgebra structure with, for all $a\in A$ 
\begin{equation}\label{letter_grouplike}
\Delta_h(a)=a\otimes a\ ;\ \ep_{aug}(a)=1	
\end{equation}
this bialgebra $(\ncp{k}{A},\ast,\Delta_h,\ep_{aug})$ is a substructure of the bialgebra of the free group.

\bs\noindent
\subsection*{Algebra of polynomials (commutative polynomials)}

We continue with the same alphabet $A$, but this time, we take as algebra $k[A]$. The construction is similar but the monomials, instead of words, are all the commutative products of letters i.e. 
$a_1^{\al_1}a_2^{\al_2}\cdots a_n^{\al_n}$ with $n$ arbitrary and $\al_i\in \N$. Denoting $\MON(A)$ the monoid of these monomials (comprising, as neutral, the empty one) and with $\Delta_s(a)=a\otimes 1+1\otimes a$, $\ep(P)=P(1_{\MON(A)})$, one can again check that $(k[A],\ast,\Delta_s,\ep)$ is a bialgebra.

\bs\noindent
\subsection*{Algebra of partially commutative polynomials}

For the detailed construction of a partially commutative monoid, the reader is referred to \cite{CF,DK}. These monoids generalize both the free and free commutative monoids. To a given graph (non-oriented and without loops) $\vartheta\subset A\times A$, one can asssociate the monoid presented by generators and relations (see basic structures \mref{basic_structures} and diagram \mref{gen_rel_universal})
\begin{equation}
	M(A,\vartheta)=\langle A;(xy=yx)_{(x,y)\in \vartheta}\rangle_{\bf Mon}\ .
\end{equation}
This is exactly the monoid obtained as a quotient structure of the free monoid ($A^*$) by the smallest equivalence compatible with products (a congruence\footnote{See exercise \mref{quotient_monoid} and the presentation of monoids ``by generators and relations in paragraph ''\mref{basic_structures}.}) which contains the pairs $(xy,yx)_{(x,y)\in \vartheta}$. A geometric model of this monoid using {\em pieces} was developped by X. Viennot \cite{Vi} where pieces are located on ``positions'' (drawn on a plane) two pieces ``commute'' iff they are on positions which do not intersect (see fig 1 below).  


\vspace{0.5cm}
\ifx\JPicScale\undefined\def\JPicScale{1}\fi
\psset{unit=\JPicScale mm}
\psset{linewidth=0.3,dotsep=1,hatchwidth=0.3,hatchsep=1.5,shadowsize=1,dimen=middle}
\psset{dotsize=0.7 2.5,dotscale=1 1,fillcolor=black}
\psset{arrowsize=1 2,arrowlength=1,arrowinset=0.25,tbarsize=0.7 5,bracketlength=0.15,rbracketlength=0.15}
\begin{pspicture}(0,0)(110,50)
\pspolygon[fillstyle=vlines](40,10)(70,10)(70,15)(40,15)
\pspolygon[fillstyle=vlines](60,15)(90,15)(90,20)(60,20)
\pspolygon[fillstyle=vlines](80,10)(110,10)(110,15)(80,15)
\pspolygon[fillstyle=vlines](80,20)(110,20)(110,25)(80,25)
\pspolygon[fillstyle=vlines](80,25)(110,25)(110,30)(80,30)
\pspolygon[fillstyle=vlines](80,30)(110,30)(110,35)(80,35)
\pspolygon[fillstyle=vlines](60,35)(90,35)(90,40)(60,40)
\pspolygon[fillstyle=vlines](40,40)(70,40)(70,45)(40,45)
\pspolygon[fillstyle=vlines](40,45)(70,45)(70,50)(40,50)
\pspolygon[fillstyle=vlines](80,40)(110,40)(110,45)(80,45)
\end{pspicture}

\rput{0}(55,09){\Large\bf $a$}
\rput{0}(75,10){\Large\bf $b$}
\rput{0}(95,09){\Large\bf $c$}

\ss
{\it {\bf Fig 1}\pointir Representation of a partially commutative monomial by ``heaps of pieces'' \cite{Vi}. Here ``$a$'' commutes with ``$c$'' (they are on non-concurrent positions) and ``$b$'' commutes with no other piece. The monomial represented can be written by several words as $acbc^3ba^2c=cabc^3baca=cabc^3bca^2$.}

\bs
The partially commutative algebra $\ncp{k}{A,\vartheta}$ is the algebra $k[M(A,\vartheta)]$ \cite{DK}. Again, one can check that 
$(\ncp{k}{A,\vartheta},\ast,\Delta_s,\ep)$ (constructed as above) is a bialgebra.  

\bs\noindent
\subsection*{Algebra of a group}

Let $G$ be a group. The algebra under consideration is $k[G]$. We define, for $g\in G$, $\Delta(g)=g\otimes g$ and $\ep(g)=1$, then, one can check that $(k[G],.,\Delta,\ep)$ is a bialgebra.

\section{Antipode and the problem of duals}

Each vector space $V$ comes with its dual\footnote{In general, for two $k$-vector spaces $V$ and $W$, $Hom_k(V,W)$ is the set of all linear mappings $V\mapsto W$.}
\begin{eqnarray}
V^*=Hom_\mathbb{C}(V,\mathbb{C}).
\end{eqnarray}
The spaces $V^*$ and $V$ are in duality by
\begin{eqnarray}
\langle p,\psi\rangle=p(\psi)
\end{eqnarray}

Now, if one has a representation (on the left) of $\mathcal{A}$ on $V$, one gets a representation on the right on $V^*$ by 
\begin{eqnarray}
\langle p.a,\psi\rangle=\langle p,a.\psi\rangle
\end{eqnarray}
If we want to have the action of $\A$ on the left again, one should use an anti-morphism $\alpha:\mathcal{A}\longrightarrow\mathcal{A}$ that is $\al\in End_k(\A)$ such that, for all $x,y\in \A$
\begin{eqnarray}
\alpha(xy)=\alpha(y)\alpha(x).
\end{eqnarray}
In the case of groups, $g\longrightarrow g^{-1}$ does the job; in the case of Lie algebras $g\longrightarrow-g$ (extended by reverse products to the enveloping algebra) works.\\ 
On the other hand, in the classical textbooks, the discussion of ``complete reductibility'' goes with the existence of an ``invariant'' scalar product $\phi : V\times V\mapsto k$ on a space $V$.\\
For a group $G$, this reads
\begin{equation}\label{group_inv}
(\forall g\in G)(\forall x,y\in V)\Big(\phi(g.x,g.y)=\phi(x,y)\Big)
\end{equation}
(think of unitary representations).\\
For a Lie algebra $\mathfrak{G}$, this reads
\begin{equation}\label{Lie_inv}
(\forall g\in \mathfrak{G})(\forall x,y\in V)\Big(\phi(g.x,y)+\phi(x,g.y)=0\Big)
\end{equation}
(think of the Killing form).\\
Now, linearizing the situation by $\Phi(x\otimes y)=\phi(x,y)$ and remembering our unit representation, one can rephrase \mref{group_inv} and \mref{Lie_inv} in 
\begin{equation}\label{phi_inv0}
(\forall a\in \A)(\forall x,y\in V)\Big(\Phi(\sum_{(1)(2)}a_{(1)}.x\otimes a_{(2)}.y)=\ep(a)\Phi(x\otimes y)\Big).
\end{equation}
Likewise, we will say that a bilinear form $\Phi : U\otimes V\mapsto k$ is invariant is it satisfies 
\begin{equation}\label{phi_inv}
(\forall a\in \A)(\forall x\in U,y\in V)\Big(\Phi(\sum_{(1)(2)}a_{(1)}.x\otimes a_{(2)}.y)=\ep(a)\Phi(x\otimes y)\Big).
\end{equation}
which means that $\Phi$ is $\A$ linear for the structures being given respectively by $\rho_1\sqtimes \rho_2$ on $U\otimes_k V$ and $\ep$ on $k$.  

Now, suppose that we have constructed a representation of a certain algebra $\A$ on $U^*$ by means of an antimorphism $\al: \A\mapsto \A$. To require that the natural contraction $c : U^*\otimes U\mapsto k$ be ``invariant'' means that 

\begin{eqnarray}\label{antipode0}
\hspace{-1.8cm}(\forall a\in \A)(\forall f\in U^*,y\in U)\Big(c(\sum_{(1)(2)}a_{(1)}.f\otimes a_{(2)}.y)=\ep(a)c(f\otimes y)=
\ep(a)f(y)\Big);\cr
(\forall a\in \A)(\forall f\in U^*,y\in U)\Big(\sum_{(1)(2)}f(\al(a_{(1)})a_{(2)}.y)=\ep(a)f(y)\Big).
\end{eqnarray}
It is easy to check (taking a basis $(e_i)_{i\in I}$ and its dual family $(e_i^*)_{i\in I}$ for example) that \mref{antipode0} is equivalent to 
\begin{equation}\label{antipode1}
(\forall a\in \A)\Big(\rho_U(\sum_{(1)(2)} \al(a_{(1)})a_{(2)})=\ep(a)Id_U\Big).	
\end{equation}
Likewise, taking the natural contraction $c : U\otimes U^*\mapsto k$, one gets 
\begin{equation}\label{antipode2}
(\forall a\in \A)\Big(\rho_U(\sum_{(1)(2)} a_{(1)}\al(a_{(2)}))=\ep(a)Id_U\Big).	
\end{equation}
Taking $U=\A$ and for $\rho_U$ the left regular representation, one gets 
\begin{equation}\label{antipode3}
(\forall a\in \A)\Big(\sum_{(1)(2)} a_{(1)}\al(a_{(2)})=\sum_{(1)(2)} \al(a_{(1)})a_{(2)})=\ep(a)\Big).	
\end{equation}
 
Motivated by the preceding discussion, one can make the following definition.
\begin{definition} Let $(\mathcal{A},\cdot,1_\mathcal{A},\Delta,\epsilon)$ be a bialgebra. 
A linear mapping $\al : \A\mapsto \A$ is called an antipode for $\A$, if for all $a\in\mathcal{A}$
\begin{eqnarray}\label{antipode}
\sum_{(1)(2)}\alpha(a_{(1)})a_{(2)}=\sum_{(1)(2)}a_{(1)}\alpha(a_{(2)})=1_\mathcal{A}\epsilon(a).
\end{eqnarray}
\end{definition}  
One can prove (see exercise \mref{convolution}), that this means that $\al$ is the inverse of $Id_\A$ for a certain product of an algebra (AAU) on $End_k(\A)$ and this implies (see exercise \mref{convolution}) that 
 
\begin{enumerate}

\item{If $\alpha$ exists (as a solution of Eq.\mref{antipode}), it is unique.}

\item{If $\alpha$ exists, it is an antimorphism.}

\end{enumerate}

\begin{definition}: (Hopf Algebra)
$(\mathcal{A},\cdot,1_\mathcal{A},\Delta,\epsilon ,S)$ is said to be a Hopf algebra iff
\begin{enumerate}
\item[(1)]{$\mathcal{B}=(\mathcal{A},\cdot ,1_\mathcal{A},\Delta,\epsilon )$ is a bialgebra,}

\item[(2)]{$S$ is an antipode (then unique) for $\mathcal{B}$}
\end{enumerate}
\end{definition}

In many combinatorial cases (see exercise on local finiteness \mref{locally_finite}), one can compute the antipode by
\begin{eqnarray}\label{antipode_by_series}
\alpha(d)=\sum_{k=0}^\infty(-1)^{k+1} (I^+)^{(*k)} (d)).
\end{eqnarray}
where $I^+$ is the projection on $\B^+=ker(\ep)$ such that $I^+(1_\B)=0$.

\section{Hopf algebras and partition functions}

\subsection{Partition Function Integrand} 

Consider the Partition Function $Z$ of a Quantum Statistical Mechanical System

\begin{equation}
Z = Tr\ exp(-\beta H)
\end{equation} 
whose hamiltonian is $H$ ($\beta\equiv 1/kT$, $k= \textrm{Boltzmann's constant}\ T=\textrm{absolute temperature}$). We may evaluate the trace over any complete set of states; we choose the (over-)complete set of coherent states
\begin{equation}
|z\rangle = e^{-|z|^2/2}\sum_{n} (z^n / \sqrt{n!})a^{+n}|0\rangle 
\end{equation}
where $a^+$ is the boson creation operator satisfying $[a,a^+] = 1$ and for 
which the completeness or resolution of unity\footnote{Sometimes physicists write $d^2z$ to emphasize that the integral is two dimensional (over $\R$) but here, the l. h. s. of \mref{res_unit} is the integration of the operator valued function $z\ra |z\rangle\langle z|$ - see \cite{B_Int1} Chap. III Paragraph 3 - w.r.t. the Haar mesure of $\C$ which is $dz$.} 
property is 
\begin{equation}\label{res_unit}
\frac{1}{\pi} \int dz |z\rangle\langle z| = I \equiv \int d�(z) |z\rangle\langle z|.  	
\end{equation}
\newpage 

The simplest, and generic, example is the free single boson hamiltonian 
$H =\ep a^+ a$ for which the appropriate trace calculation is 
\begin{eqnarray}
Z = \frac{1}{\pi} \int dz \langle z| exp(-\beta a^+ a) |z\rangle =\cr 
= \frac{1}{\pi}\int dz \langle z| : exp(a^+ a(e^{-\beta\ep} - 1) : |z\rangle\ .
\end{eqnarray}
Here we have used the following well known relation for the 
forgetful normal ordering operator $: f(a, a^+ ) :$ which means ``normally 
order the creation and annihilation operators in $f$ forgetting the commutation relation 
$[a, a^+] = 1$''\footnote{Of course, this procedure may alter the value of the operator to which it is applied.}. 
We may write the Partition Function in general as 
\begin{equation}\label{free_boson}
Z(x) = \int F (x, z) d�(z)	
\end{equation}
thereby defining the Partition Function Integrand (PFI) $F(x, z)$. We 
have explicitly written the dependence on $x= -\beta$, the inverse temperature, and $\ep$, the energy scale in the hamiltonian. 

\subsection{Combinatorial aspects: Bell numbers} 

The generic free boson example Eq. \mref{free_boson} above may be rewritten to show 
the connection with certain well known combinatorial numbers. Writing $y = |z|^2$ and $x = -\beta\ep$, Eq.\mref{free_boson} becomes 
\begin{equation}
Z = \int_0^\infty dy\ exp ( y(e^x - 1) )\ .  	
\end{equation}
This is an integral over the classical exponential generating function 
for the Bell polynomials 

\begin{equation}
exp(y(e^x - 1)) = \sum_{n=0}^\infty B_n(y) \frac{x^n}{n!} 
\end{equation}

where the Bell number is $B_n(1) = B(n)$, the number of ways of putting 
$n$ different objects into $n$ identical containers (some may be left empty). 
Related to the Bell numbers are the Stirling numbers of the second kind 
$S(n, k)$, which are defined as the number of ways of putting $n$ different 
objects into $k$ identical containers, leaving none empty. From the 
definition we have $B(n) = \sum^n_{k=1} S(n, k)$ (for $n\not=0$). 
The foregoing gives a combinatorial interpretation of the partition function 
integrand $F(x, y)$ as the exponential generating function of the Bell polynomials. 

\subsection{Graphs}
 
We now give a graphical representation of the Bell numbers. Consider 
labelled lines which emanate from a white dot, the origin, and finish on 
a black dot, the vertex. We shall allow only one line from each white 
dot but impose no limit on the number of lines ending on a black 
dot. Clearly this simulates the definition of $S(n, k)$ and $B(n)$, with 
the white dots playing the role of the distinguishable objects, whence 
the lines are labelled, and the black dots that of the indistinguishable 
containers. The identification of the graphs for 1,2 and 3 lines is given 
in Figure 2. We have concentrated on the Bell number sequence and 
its associated graphs since, as we shall show, there is a sense in which 
this sequence of graphs is generic. That is, we can represent any 
combinatorial sequence by the same sequence of graphs as in the Figure 2

\bs
\def\JPicScale{0.7}
\psset{unit=\JPicScale mm}
\psset{linewidth=0.3,dotsep=1,hatchwidth=0.3,hatchsep=1.5,shadowsize=1,dimen=middle}
\psset{dotsize=0.7 2.5,dotscale=1 1,fillcolor=black}
\psset{arrowsize=1 2,arrowlength=1,arrowinset=0.25,tbarsize=0.7 5,bracketlength=0.15,rbracketlength=0.15}
\begin{pspicture}(0,0)(215.26,85)
\rput{0}(130.08,50.01){\psellipse[linewidth=0.75,fillstyle=solid](0,0)(2.5,2.5)}
\psline[linewidth=0.75](110.26,77.24)(127.89,51.71)
\psline[linewidth=0.75](130.13,52.63)(130.13,76.97)
\psline[linewidth=0.75](132.5,51.97)(150.26,77.24)
\pspolygon[linewidth=0.75](147.66,82.52)(152.69,82.52)(152.69,77.52)(147.66,77.52)
\pspolygon[linewidth=0.75](127.65,82.52)(132.68,82.52)(132.68,77.52)(127.65,77.52)
\pspolygon[linewidth=0.75](107.48,82.63)(112.51,82.63)(112.51,77.63)(107.48,77.63)
\rput{0}(10.08,50.01){\psellipse[linewidth=0.75,fillstyle=solid](0,0)(2.5,2.5)}
\psline[linewidth=0.75](10.13,52.63)(10.13,76.97)
\pspolygon[linewidth=0.75](7.65,82.52)(12.68,82.52)(12.68,77.52)(7.65,77.52)
\rput{0}(40.08,50.01){\psellipse[linewidth=0.75,fillstyle=solid](0,0)(2.5,2.5)}
\psline[linewidth=0.75](30.26,77.24)(38.42,52.24)
\psline[linewidth=0.75](41.71,52.37)(50.13,76.97)
\pspolygon[linewidth=0.75](47.65,82.52)(52.68,82.52)(52.68,77.52)(47.65,77.52)
\pspolygon[linewidth=0.75](27.48,82.63)(32.51,82.63)(32.51,77.63)(27.48,77.63)
\rput{0}(69.94,49.62){\psellipse[linewidth=0.75,fillstyle=solid](0,0)(2.5,2.5)}
\psline[linewidth=0.75](70,52.24)(70,76.58)
\pspolygon[linewidth=0.75](67.52,82.13)(72.55,82.13)(72.55,77.13)(67.52,77.13)
\rput{0}(79.94,49.62){\psellipse[linewidth=0.75,fillstyle=solid](0,0)(2.5,2.5)}
\psline[linewidth=0.75](80,52.24)(80,76.58)
\pspolygon[linewidth=0.75](77.52,82.13)(82.55,82.13)(82.55,77.13)(77.52,77.13)
\pspolygon[](5,85)(15,85)(15,40)(5,40)
\pspolygon[](25,85)(55,84.87)(55,39.87)(25,40)
\pspolygon[](60,85)(90,84.87)(90,39.87)(60,40)
\pspolygon[](105,85)(155.26,84.87)(155.26,39.87)(105,40)
\rput{0}(180.34,49.75){\psellipse[linewidth=0.75,fillstyle=solid](0,0)(2.5,2.5)}
\psline[linewidth=0.75](180.39,52.37)(180.39,76.71)
\pspolygon[linewidth=0.75](177.91,82.26)(182.94,82.26)(182.94,77.26)(177.91,77.26)
\rput{0}(200.34,49.75){\psellipse[linewidth=0.75,fillstyle=solid](0,0)(2.5,2.5)}
\psline[linewidth=0.75](190.53,76.97)(198.68,51.97)
\psline[linewidth=0.75](201.97,52.11)(210.39,76.71)
\pspolygon[linewidth=0.75](207.91,82.26)(212.94,82.26)(212.94,77.26)(207.91,77.26)
\pspolygon[linewidth=0.75](187.74,82.37)(192.77,82.37)(192.77,77.37)(187.74,77.37)
\pspolygon[](175,84.61)(215.26,84.61)(215.26,39.61)(175,39.61)
\end{pspicture}

\vspace{-2cm}
{\small{\bf Fig 2}\pointir \it Graphs for B(n), n = 1, 2, 3}.

\bs 
with suitable vertex multipliers (denoted by the $V$ terms in the same 
figure). Consider a general partition function 
\begin{equation}\label{part_func8}
Z = Tr\ exp(-\beta H)	
\end{equation}
 
where the Hamiltonian is given by $H = \ep w(a, a^+)$, with $w$ a string 
($=$ sum of products of positive powers) of boson creation and annihilation 
operators. The partition function integrand $F$ for which we seek to 
give a graphical expansion, is 
\begin{equation}
Z(x) = \int F (x, z) d�(z)	
\end{equation}
where 
\begin{eqnarray}
F (x, z) &=& \langle z| exp(xw)|z\rangle =\hspace{1cm} (x = -\beta\ep)\cr 
         &=& \sum_{n=0}^\infty  \langle z|w^n|z\rangle \frac{x^n}{n!}\cr 
         &=& \sum_{n=0}^\infty W_n (z) \frac{x^n}{n!} \cr
         &=& exp\Big(\sum_{n=1}^\infty  V_n (z) \frac{x^n}{n!} \Big)
\end{eqnarray}

with obvious definitions of $W_n$ and $V_n$. The sequences $\{W_n\}$ and $\{V_n\}$ 
may each be recursively obtained from the other. This relates 
the sequence of multipliers $\{V_n\}$ of Figure 2 to the Hamiltonian of 
Eq. \mref{part_func8}. The lower limit $1$ in the $V_n$ summation is a consequence of the 
normalization of the coherent state $|z\rangle$. 

\subsection{The Hopf Algebra $\mathcal{L}_{\rm Bell}$} 
We briefly describe the Hopf algebra $\LB$ which the diagrams of Figure 2 define.\\ 
1. Each distinct diagram is an individual basis element of $\LB$ ; thus 
the dimension is infinite. (Visualise each diagram in a ``box''.) 
The sum of two diagrams is simply the two boxes containing the 
diagrams. Scalar multiples are formal; for example, they may be 
provided by the V coefficients. Precisely, as a vector space, $\LB$ is the space
freely generated by the diagrams of Figure 2 (see APPENDIX : Function Spaces).\\  
2. The identity element e is the empty diagram (an empty box).\\ 
3. Multiplication is the juxtaposition of two diagrams within the 
same ``box''. $\LB$ is generated by the connected diagrams; this 
is a consequence of the Connected Graph Theorem \cite{FU}. Since we 
have not here specified an order for the juxtaposition, multipli� 
cation is commutative.\\ 
4. The comultiplication $\Delta : \LB\mapsto \LB\otimes \LB$ is defined by\\ 
\begin{eqnarray}
\Delta(e) = e \otimes e\textrm{  (unit $e$, the empty box)}\cr	
\Delta(x) = x \otimes e + e \otimes x \textrm{ (generator $x$)}\cr
\Delta(AB) = \Delta(A)\Delta(B) \textrm{ otherwise}\cr
\textrm{so that }\Delta \textrm{ is an algebra homomorphism.} 
\end{eqnarray}
 
\section{The case of two modes}

Let us consider an hamiltonian on two modes $H(a,a^+,b,b^+)$, with 
\begin{eqnarray}
[a,a^+]=1\cr
[b,b^+]=1\cr
[a^{\ep_1},b^{\ep_2}]=0,\ \ep_i \textrm{ being $+$ or empty (4 relations)}
\end{eqnarray}

Suppose that one can express $H$ as 
\begin{equation}
H(a,a^+,b,b^+)=H_1(a,a^+)+H_2(b,b^+).	
\end{equation}

and that, $exp(\lambda H_1)$ and $exp(\lambda H_2)$ ($\lambda=-\beta$) are solved i.e. 
that we have expressions  
\begin{equation}\hspace{-2cm}
F(\lambda)=exp(\lambda H_1)=\sum_{n=0}^\infty	\frac{\lambda^n}{n!} H_1^{(n)}(a,a^+)\ ;\  
G(\lambda)=exp(\lambda H_2)=\sum_{n=0}^\infty	\frac{\lambda^n}{n!} H_2^{(n)}(b,b^+)
\end{equation}

It is not difficult to check that 
\begin{equation}
	exp(\lambda H)=exp\Big(\lambda (H_1+H_2)\Big)=F(\lambda \frac{d}{dx})G(x)\Big|_{x=0}
\end{equation}

This leads us to define, in general, the ``Hadamard exponential product''. Let 

\begin{eqnarray}
F(z)=\sum_{n\geq 0} a_n\frac{z^n}{n!},\ G(z)=\sum_{n\geq 0}
b_n\frac{z^n}{n!} 
\end{eqnarray}

and define their product (the ``Hadamard exponential product'') by

\begin{eqnarray}
\mathcal{H}(F,G):=\sum_{n\geq 0}
a_nb_n\frac{z^n}{n!}=\mathcal{H}(F,G)=\left.F\left(z\frac{d}{dx}\right)G(x)\right|_{x=0}.
\end{eqnarray}  

When $F(0)$ and $G(0)$ are
not zero one can normalize the functions in this bilinear product so  that $F(0)=G(0)=1$.
We would like to obtain compact and generic formulas. If we
write the functions as

\begin{eqnarray}
F(z)=\exp\left(\sum_{n=1}^\infty L_n\frac{z^n}{n!}\right) \textrm{ and } G(z)=\exp\left(\sum_{n=1}^\infty V_n\frac{z^n}{n!}\right).
\end{eqnarray}
that is, as free exponentials, then by using   Bell polynomials
in the sets of variables $\L, \V$ (see \cite{GOF4,OPG} for details), we obtain

\begin{eqnarray}
\mathcal{H}(F,G)=\sum_{n\geq 0} \frac{z^n}{n!} \sum_{P_1,P_2\in
UP_n} \L^{Type(P_1)}\V^{Type(P_2)}
\end{eqnarray}
where $UP_n$ is the set of unordered partitions of $[1\cdots n]$. An
unordered partition $P$ of a set $X$ is a collection of (nonempty) subsets of $X$, 
mutually disjoint and covering $X$ (i.e. the union of all the subsets is $X$, see 
\cite{GOF12} for details).

The type of $P\in UP_n$ (denoted above by $Type(P)$) is the multi-index
$(\al_i)_{i\in \N^+}$ such that $\al_k$ is the number of $k$-blocks,
that is the number of members of $P$ with cardinality $k$.

\ss
At this point  the formula entangles and the diagrams of the theory arise.\\
Note particularly that
\begin{itemize}
    \item the monomial $\L^{Type(P_1)}\V^{Type(P_2)}$ needs much less information than that which is contained in the individual partitions $P_1,\ P_2$ (for example, one can relabel the elements without changing the monomial),
    \item two partitions have an incidence matrix {\it from which it is still possible to recover the types of the partitions.}
\end{itemize}

\ss The construction now proceeds as follows.
\begin{enumerate}
    \item Take two unordered partitions of $[1\cdots n]$, say $P_1,P_2$.
    \item Write down their incidence matrix $\left(\card(Y\cap Z)\right)_{(Y,Z)\in P_1\times P_2}$.
    \item Construct the diagram representing the multiplicities of the incidence matrix : for each block of $P_1$ draw a black spot (resp. for each block of $P_2$ draw a white spot).
    \item Draw lines between the black spot $Y\in P_1$ and the white spot $Z\in P_2$; there are  $\card(Y\cap Z)$ such.
    \item Remove the information of the blocks $Y,Z,\cdots$.
\end{enumerate}

In so doing, one obtains a bipartite graph with $p$ ($=\card(P_1)$)
black spots, $q$ ($=\card(P_2)$) white spots, no isolated vertex and
integer multiplicities. We denote the set of such diagrams by $\diag$.

\bs
\def\JPicScale{1}
\unitlength \JPicScale mm
\begin{picture}(122.5,87.5)(0,5)
\linethickness{0.75mm}
\multiput(69.34,49.61)(0.12,0.37){76}{\line(0,1){0.37}}
\linethickness{0.75mm}
\multiput(72.23,52.11)(0.12,0.37){67}{\line(0,1){0.37}}
\linethickness{0.75mm}
\multiput(82.76,78.82)(0.12,-0.57){48}{\line(0,-1){0.57}}

\put(60,80){\circle{5}}
\put(80.13,79.87){\circle{5}}
\put(100.13,79.87){\circle{5}}
\put(120,80){\circle{5}}

\put(70.2,49.88){\circle*{5}}
\put(90.3,49.56){\circle*{5}}
\put(110,50){\circle*{5}}

\put(56,86){$\{1\}$}
\put(70,86){$\{2,3,4\}$}
\put(88,86){$\{5,6,7,8,9\}$}
\put(115,86){$\{10,11\}$}

\put(60.2,41.88){$\{2,3,5\}$}
\put(77.3,41.88){$\{1,4,6,7,8\}$}
\put(104,41.88){$\{9,10,11\}$}

\linethickness{0.75mm}
\multiput(71.97,49.74)(0.12,0.14){212}{\line(0,1){0.14}}

\linethickness{0.75mm}
\multiput(102.76,79.47)(0.12,-0.48){57}{\line(0,-1){0.48}}

\linethickness{0.75mm}
\multiput(92.76,49.61)(0.12,0.36){81}{\line(0,1){0.36}}
\linethickness{0.75mm}
\multiput(90.26,48.29)(0.12,0.37){78}{\line(0,1){0.37}}
\linethickness{0.75mm}
\multiput(88.29,49.08)(0.12,0.36){81}{\line(0,1){0.36}}
\linethickness{0.75mm}
\multiput(111.97,50.66)(0.12,0.37){72}{\line(0,1){0.37}}
\linethickness{0.75mm}
\multiput(109.74,51.05)(0.12,0.37){73}{\line(0,1){0.37}}
\linethickness{0.75mm}
\multiput(61.71,78.03)(0.12,-0.13){216}{\line(0,-1){0.13}}
\end{picture}

\vspace{-2cm}

{\small{\bf Fig 3}\pointir \it Diagram from $P_1,\ P_2$ (set partitions of $[1\cdots 11]$).\\ 
$P_1=\left\{\{2,3,5\},\{1,4,6,7,8\},\{9,10,11\}\right\}$ and  $P_2=\left\{\{1\},\{2,3,4\},\{5,6,7,8,9\},\{10,11\}\right\}$ (respectively black spots for $P_1$ and white spots for $P_2$).\\ 
The incidence matrix corresponding to the diagram (as drawn) or these partitions is 
${\pmatrix{0 & 2 & 1 & 0\cr 1 & 1 & 3 & 0\cr 0 & 0 & 1 & 2}}$. But, due to the fact that the defining partitions are unordered, one can permute the spots (black and white, between themselves) and, so, the lines and columns of this matrix can be permuted. Thus, the diagram could be represented by the matrix ${\pmatrix{0 & 0 & 1 & 2\cr 0 &  2 & 1 & 0\cr 1 & 0 & 3 & 1}}$ as well.}


\bs The product formula now reads

\begin{eqnarray}
\mathcal{H}(F,G)=\sum_{n\geq 0} \frac{z^n}{n!} \sum_{d\in diag\atop
|d|=n} mult(d)\L^{\al(d)}\V^{\be(d)}
\end{eqnarray}
where $\al(d)$ (resp. $\be(d)$) is the ``white spot type'' (resp.
the ``black spot type'') i.e. the multi-index $(\al_i)_{i\in \N^+}$
(resp. $(\be_i)_{i\in \N^+}$) such that $\al_i$ (resp. $\be_i$) is
the number of white spots (resp. black spots) of degree $i$ ($i$
lines connected to the spot) and $mult(d)$ is the number of pairs of
unordered partitions of $[1\cdots |d|]$ (here
$|d|=|\al(d)|=|\be(d)|$ is the number of lines of $d$) with
associated diagram $d$.

\begin{no_te} The diagrams as well as the product formula were introduced in \cite{BBM}. 
\end{no_te}

\subsection{Diagrams}\label{diagrams}

One can design a (graphically) natural multiplicative structure
on $\diag$ such that the arrow
\begin{equation}
    d \mapsto \L^{\al(d)}\V^{\be(d)}.
\end{equation}
be a morphism.
This is provided by the concatenation of the diagrams (the result, i.e. the diagram obtained
in placing $d_2$ at the right of $d_1$, will be denoted by
$[d_1|d_2]_D$). One must check that this product is compatible with
the equivalence of the permutation of white and black spots among themselves, which is rather straightforward (see \cite{GOF4,GOF12}). We have

\begin{proposition}\cite{GOF12} Let $\diag$ be the set of diagrams (including the empty one).\\
i) The law $(d_1,d_2)\mapsto [d_1|d_2]_D$ endows $\diag$ with the structure of a commutative monoid with the empty diagram as neutral element(this diagram will, therefore, be denoted by $1_{\diag}$).\\
ii) The arrow $d \mapsto \L^{\al(d)}\V^{\be(d)}$ is a morphism of
monoids, the codomain of this arrow being the monoid of
(commutative) monomials in the alphabet $\L\cup \V$ i.e.
\begin{equation}
\hspace{-2.4cm}\mathfrak{MON}(\L\cup\V)=\{\L^\al \V^\be\}_{\al,\be\in (\N^+)^{(\N)}}=
\bigcup_{n,m\geq 1}\big\{L_1^{\al_1}L_2^{\al_2}\cdots
L_n^{\al_n}V_1^{\be_1}V_2^{\be_2}\cdots
V_m^{\be_m}\big\}_{\al_i,\be_j\in \N}.	
\end{equation}
iii) The monoid $(\diag,[-|-]_D,1_{\diag})$ is a free commutative monoid. 
The set on which it is built is the set of the connected (non-empty) diagrams.
\end{proposition}

\begin{remark} The reader who is not familiar with the algebraic structure of $\MON(X)$ can find rigorous definitions in paragraph \mref{basic_structures}.
\end{remark}

We denote $\phi_{mon,diag}$ the arrow $\diag\mapsto \mathfrak{MON}(\L\cup\V)$. 

\subsection{Labelled diagrams}\label{labelled_diagrams} 

We have seen the diagrams (of $\diag$) are in one-to-one correspondence with classes of matrices as in Fig. 3.  
In order to fix one representative of this class, we have to number the black (resp. white) spots from 1 to, say $p$ (resp. $q$). Doing so, one obtains a {\em packed matrix} \cite{DHT} that is, a matrix of integers with no row nor column consisting entirely of zeroes. In this way, we define the \textit{labelled diagrams}.

\begin{definition}
	A labelled diagram of size $p\times q$ is a bi-coloured (vertices are $p$ black and $q$ white spots) graph
	
\begin{itemize}
	\item with no isolated vertex
	\item every black spot is joined to a white spot by an arbitrary quantity (but a positive integer) of lines
	\item the black (resp. white) spots are numbered from 1 to $p$ (resp. from 1 to $q$).
\end{itemize}
\end{definition} 

As in paragraph \mref{diagrams}, one can concatenate the labelled diagrams, the result, i.e. the diagram obtained
in placing $D_2$ at the right of $D_1$, will be denoted by
$[D_1|D_2]_{L}$. This time we need not  check  compatibility with classes. We have a structure of free monoid (but not commutative this time)

\begin{proposition}\cite{GOF12} Let $\ldiag$ be the set of labeled diagrams (including the empty one).\\
i) The law $(d_1,d_2)\mapsto [d_1|d_2]_L$ endows $\ldiag$ with the structure of a noncommutative monoid with the empty diagram ($p=q=0$) as neutral element (which will, therefore, be denoted by $1_{\ldiag}$).\\
ii) The arrow from $\ldiag$ to $\diag$, which implies ``forgetting the labels of the vertices'' is a morphism of monoids.\\
iii) The monoid $(\ldiag,[-|-]_L,1_{\ldiag})$ is a free
(noncommutative) monoid which is constructed on the set of irreducible diagrams which are diagrams $d\not= 1_{\ldiag}$ which cannot be written $d=[d_1|d_2]_L$ with $d_i\not=1_{\ldiag}$.
\end{proposition}

\begin{remark}
i) In a general monoid $(M,\star,1_M)$, the irreducible elements are the elements $x\neq 1_M$ such that $x=y\star z\Longrightarrow 1_M\in \{y,z\}$.\\
ii) It can happen that an irreducible of $\ldiag$ has an image in
$\diag$ which splits (i. e. is reducible), as shown by  the simple example of the {\em cross}
defined by the incidence matrix ${\pmatrix{0 & 1\cr 1 & 0}}$.
\end{remark}

\subsection{Hopf algebras $\DIAG$ and $\LDIAG$}

Let us first construct the Hopf algebra on the labelled diagrams (details can be found in \cite{GOF12}). In order to define the comultiplication, we need the notion of ``restriction of a labelled diagram''. Consider $d\in \ldiag$ of size $p\times q$. For any subset $I\subset [1..p]$, we define a labelled diagram $d[I]$ (of size $k\times l$, 
$k=\card(I)$) by taking the $k=|I|$ black spots numbered in $I$ and the edges (resp. white spots) that are connected to them. We take this subgraph and relabel  the black (resp. white) spots in increasing order.\\
The construction of the Hopf algebra $\LDIAG$ goes as follows :\\
\begin{enumerate}
	\item the algebra structure is that of algebra of the monoid $\ldiag$ so that the elements of $\LDIAG$ are
\begin{equation}\label{gen_elem}
	x=\sum_{d\in \ldiag} \al_d d
\end{equation}
(the sum is finitely supported) 
	\item the comultiplication is given, on a labelled diagram $d\in \ldiag$ of size $p\times q$, by
\begin{equation}
	\Delta_L(d)=\sum_{I+J=[1..p]} d[I]\times d[J]
\end{equation}
	\item the counit is ``taking the coefficient of the void diagram'',\\ 
	that is, for $x$ as in Eq. \mref{gen_elem}, 	
\begin{equation}
\ep_L(x)=	\al_{1_\ldiag}.
\end{equation}
\end{enumerate}
   
One can check that $(\LDIAG,[-|-]_L,1_{\ldiag},\Delta_L,\ep_L)$ is a bialgebra (for proofs see \cite{GOF12}). Now one can check that we satisfy the conditions of exercise \mref{locally_finite} question 3 and the antipode $S_L$ can be computed by formula \mref{antipode_by_series1} of the same exercise. 

\ss
We have so far constructed the Hopf algebra $(\LDIAG,[-|-]_L,1_{\ldiag},\Delta,\ep,S_L)$. 

\ss
The constructions above are compatible with the arrow 
$$
\phi_{\DIAG,\LDIAG}\ :\ \LDIAG\mapsto \DIAG
$$ 
deduced from the class-map $\phi_{\diag,\ldiag}\ :\ \ldiag\mapsto \diag$ (a diagram is a class of labelled diagrams under permutations of black and white spots among themselves). So that, one can deduce ``by taking quotients'' a structure of Hopf algebra on the algebra of $\diag$. Denoting this algebra by $\DIAG$, one has a natural  Hopf algebra structure 
$(\DIAG,[-|-]_D,1_{\diag},\Delta_D,\ep_D,S_D)$ and one can prove that this is the unique Hopf algebra structure   such that $\phi_{\DIAG,\LDIAG}$ is a morphism for the algebra and coalgebra structures.

\section{Link between $\LDIAG$ and other Hopf algebras}

\subsection{The deformed case}

One can construct a three-parameter Hopf algebra deformation of $\LDIAG$ (see \cite{GOF12}), denoted $\LDIAG(q_c,q_s,q_t)$ such that $\LDIAG(0,0,0)=\LDIAG$ and \\
$\LDIAG(1,1,1)=\MQS$ the algebra of Matrix Quasi Symmetric Functions \cite{DHT}). On the other hand, 
it was proved by L. Foissy \cite{F2,F3} that one of the planar decorated trees Hopf algebra is isomorphic $\MQS$ and even to $\LDIAG(1,q_s,t)$ for every $q_s$ and $t\in\{0,1\}$. The complete picture is given below.

\begin{center}
\includegraphics[scale=0.6]{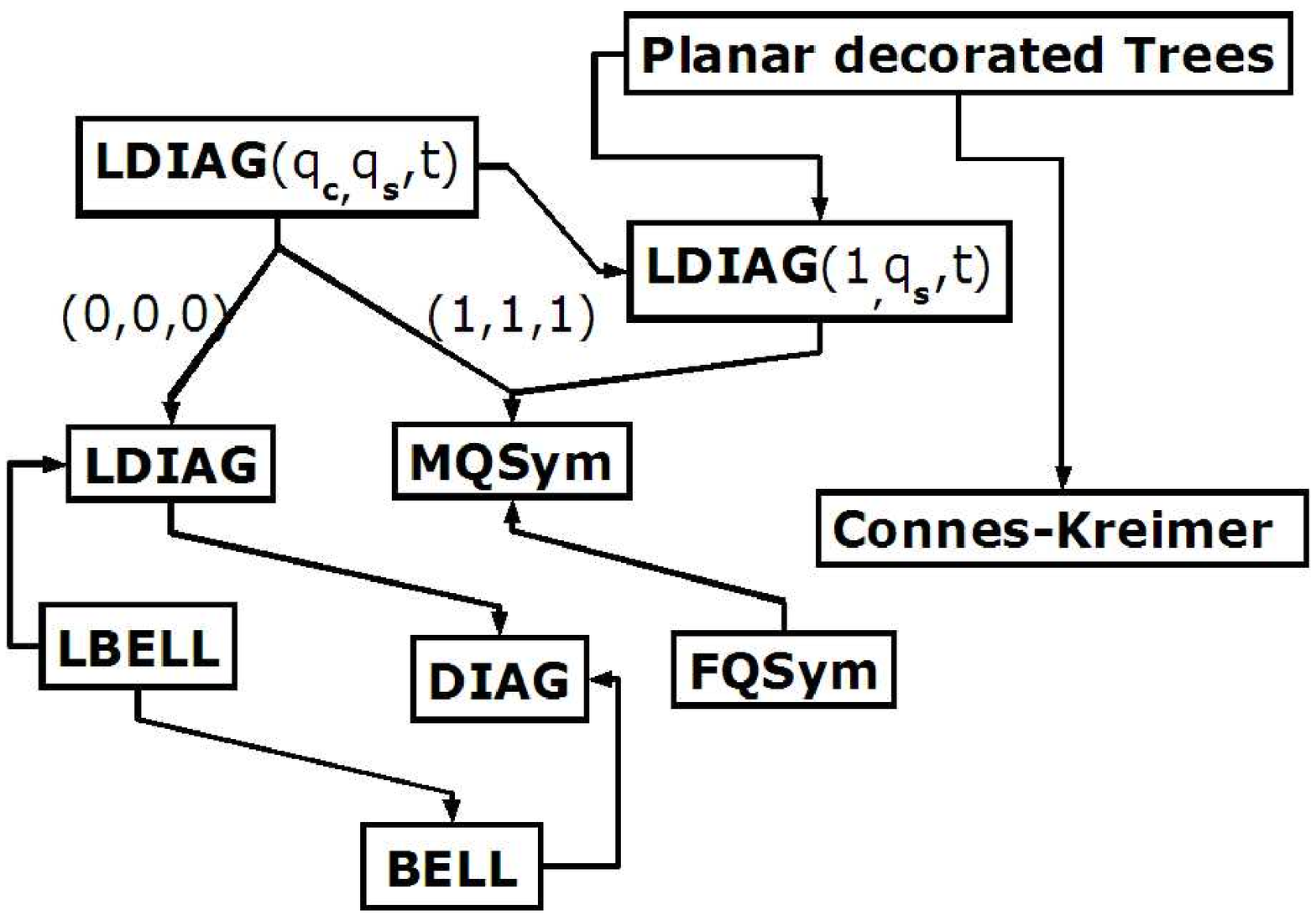}
\end{center}

\section{Duals of Hopf algebras}

The question of dualizing a Hopf algebra (i.e. endowing the dual - or a subspace of it - with a structure of Hopf algebra) is solved, in complete generality, by the machinery of Sweedler's duals. The procedure consists in taking the ``representative'' linear forms (instead of all the linear forms) and dualize w.r.t. the following table 

\begin{center}
\begin{tabular}{ccc}
comultiplication  & $\ra$ & multiplication\\
counit & $\ra$ & unit\\
multiplication & $\ra$ & comultiplication\\
unit & $\ra$ & counit\\
antipode & $\ra$ & trsnspose of the antipode.
\end{tabular}
\end{center}

In the case when the Hopf algebra is free as an algebra (which is often the case with noncommutative Hopf algebras of combinatorial physics), one can use rational expressions of Automata Theory to get a genuine calculus within this dual (see \cite{DT}).\\

\newpage  

\section{EXERCISES}

\begin{exo}\label{complex_rep_func} \textbf{Representative functions on $\mathbb{R}$ (Complex valued)}. A function $\mathbb{R}\stackrel{f}{\longrightarrow}\mathbb{C}$ is said to be representative if there exist $(f_i^{(1)})_{i=1}^n$ and $(f_i^{(2)})_{i=1}^n$ such that, for all $x,y\in\mathbb{R}$ one has
\begin{eqnarray}\label{Repr}
f(x+y)=\sum_{i=1}^nf_i^{(1)}(x)f_i^{(2)}(y).
\end{eqnarray}
1) Show that $\cos$, $\cos^2$, $\sin$, $\exp$ and $a\ra a^n$ are representative. Provide minimal sums of type Eq.(\ref{Repr}).\\
2) Show that the following are equivalent\\
i) $f$ is representative.\\
ii) There exists a group representation $(\mathbb{R},+)\stackrel{\rho}{\longrightarrow}\C^{n\times n}$, a row vector $\lambda\in\C^{1\times n}$, a column vector $\gamma\in\C^{n\times 1}$ such that $f(x)=\lambda\rho(x)\gamma$.\\
iii) $(f_t)_{t\in\mathbb{R}}$ is of finite rank in $\C^\mathbb{R}$ (here $f_t$, the shift of $f$ by $t$, is the function $x\longrightarrow f(x+t)$).

\ss
3) Show that the minimal $n$ such that formula Eq.(\ref{Repr}) holds is also the rank (over $\C$) of $(f_t)_{t\in\mathbb{R}}$.

4) a) If $f$ is continuous then $\rho$ can be chosen so and $\rho(x)=e^{xT}$ for a certain matrix $T\in\C^{n\times n}$.

b) In this case show that representative functions are linear combinations of products of polynomials and exponentials.

5) $f\in\mathbb{C}^\mathbb{R}$ is representative iff $\mathcal{R}e(f)=(f+\bar{f})/2$ and $\mathcal{I}m(f)=(f-\bar{f})/2i$ are representative in $\mathbb{R}^\mathbb{R}$.

6) Show that the set of representative functions of $\C^\R$ is a $\C$-vector space. This space will be denoted $Rep_\C(\mathbb{R})$.

7) Show that the functions $\varphi_{n,\lambda}=x^ne^{\lambda x}$ are a basis of $Rep_\mathbb{C}(\mathbb{R})\cap\mathcal{C}_o(\mathbb{R};\mathbb{C})$ (i.e. continuous complex valued representative functions on $\R$).

8) Deduce from (7) that the following statement is false:\\
``If a entire function $f:\R\mapsto \C$ is such that $(f_t)_{t\in \Z}$ is of finite rank, then it is representative''.
{\footnotesize {\bf Hint} : Consider $exp(exp(2i\pi x))$.}
\end{exo}

\begin{exo}\label{rep_func} \textbf{Representative functions in general} (see also \cite{Ab,CP}) Let $M$ be a monoid (semigroup with unit) and $k$ a field (one can first think of $k=\mathbb{R},\mathbb{C}$). For a function $M\stackrel{f}{\longrightarrow} k$ one defines the shifts:
\begin{eqnarray}
f_z:x\longrightarrow f(zx),\cr
_yf:x\longrightarrow f(xy),\cr
_yf_z:x\longrightarrow f(zxy).
\end{eqnarray}

1)a) Check the following formulas
\begin{eqnarray}\label{f1}
(f_{y_1})_{y_2}=f_{y_1y_2},\cr
_{y_2}(_{y_1}f_)=_{y_2y_1}f\cr
(_xf)_y=_x(f_y)=_xf_y.
\end{eqnarray}

As for groups, if $M$ is a monoid, a $M$-module structure on a vector space $V$ is defined by a morphism or an anti-morphism $M\mapsto End_k(V)$.\\ 
b) From Eqs.(\ref{f1}) define two canonical $M$-module structures of $k^M$.\\ 

2)a) Show that the following are equivalent

i) $(f_z)_{z\in M}$ is of finite rank in $k^M$.

ii) $(_yf)_{y\in M}$ is of finite rank in $k^M$.

iii) $(_yf_z)_{y,z\in M}$ is of finite rank in $k^M$.

iv) There exist two families $(f_i^{(1)})_{i=1}^n$ and $(f_i^{(2)})_{i=1}^n$ such that
\begin{eqnarray}\label{add_formula}
f(xy)=\sum_{i=1}^nf_i^{(1)}(x)f_i^{(2)}(y).
\end{eqnarray}

v) There exists a representation of $M$ $\rho:M\longrightarrow k^{n\times n}$, a row vector $\lambda\in k^{1\times n}$, a column vector $\gamma\in k^{n\times 1}$ such that $f(x)=\lambda\rho(x)\gamma$ for all $x\in M$.

b) Using (v) above, show that the (pointwise) product of two representative functions is representative.

\ss
One denotes $Rep_k(M)$ the space of ($k$-valued) representative functions on $M$.\\
3) a) Recall briefly why the mapping 
\begin{equation}
k^{M}\otimes k^{M}\mapsto k^{M\times M}	
\end{equation}

defined by 
$$
\sum_{i=1}^nf_i^{(1)}\otimes f_i^{(2)}\ra \Big((x,y)\ra \sum_{i=1}^nf_i^{(1)}(x) f_i^{(2)}(y)\Big)
$$ 
is injective.\\
b) Show that, if $n$ is minimal in \mref{add_formula}, the families of functions $f_i^{(1)}$ and $f_i^{(2)}$ are 
representative and that the mapping $Rep_k(M)\mapsto Rep_k(M)\otimes Rep_k(M)$ defined by
\begin{equation}
f\ra 	\sum_{i=1}^nf_i^{(1)}\otimes f_i^{(2)}
\end{equation}
defines a structure of coassociative coalgebra on $Rep_k(M)$ Has it a co-unit ?
\\
We denote by $\Delta$ the compultiplication contructed above.\\
4) Show that $(Rep_k(M),.,\mathbf{1},\Delta,\ep_{1_M})$ (with $\mathbf{1}$ the constant-valued function equal to $1$ on $M$ and $\ep$ the Dirac measure $f\ra f(1_M)$) is a bialgebra.   
\end{exo}

\begin{exo}\label{dissipation} (Example of a monoid coming from dissipation theory \cite{S1}).\\
Let $(\M_n,\times)$ be the monoid of $n\times n$ complex square matrices endowed with the usual product 
$$
(\M_n,.)=(\C^{n\times n},\times)
$$
We define a {\rm state} (Von Neumann) as a positive semi-definite hermitian matrix of trace one ($=1$) i.e. a matrix $\rho$ such that 
\begin{enumerate}
	\item $\rho=\rho^*$
	\item $(\forall x\in \C^{n\times 1})(x^*\rho x\geq 0)$
	\item $Tr(\rho)=1$.
\end{enumerate}
The set of such states will be denoted by $\St_n$.\\ 
1) (Structure) a) Show that $\St_n$ is convex.\\ 
b) Show that $\St_n$ is compact.\\ 
{\footnotesize {\bf Hint} : Consider the set of possible spectra i.e. the simplex 
$$
S=\{(\lambda_1,\lambda_2,\cdots ,\lambda_n)\in (\R^+)^n | \sum_{i=1}^n \lambda_i=1\}
$$
and show that $\St_n$ is the range of the continuous mapping $\phi : U(n)\times S\mapsto \M_n$ given by the formula  
\begin{eqnarray}
\phi(u,s)= u d u^* \ ;\ \textrm{with } 
d=
\left(\begin{array}{cccc}
\lambda_1 &         0 & \cdots & 0\\
0         & \lambda_2 & \cdots & 0\\
          &           & \ddots & \\
0         & \cdots    & 0      &\lambda_n
\end{array}\right)
\end{eqnarray}
and where $s=(\lambda_1,\lambda_2,\cdots ,\lambda_n)$.\\
} 
c) Show that the extremal elements \cite{B_TVS} of the compact $\St_n$ is the set of orthogonal projections of rank one and that this set is connected by arcs.\\
2) (KS condition) We say that a finite family $(k_i)_{i\in I}$ ($I$ is finite) of elements in $\M_n$ fulfils the KS condition iff $\sum_{i\in I} k_i^*k_i=I_n$ ($I_n$ is the unity matrix, the unit of the monoid $\M_n$).\\ 
On the other hand, given two finite families $A=(k_i)_{i\in I}$ and $B=(l_j)_{j\in J}$, we define $A\ast B$ as the family
\begin{equation}
	A\ast B=(k_il_j)_{(i,j)\in I\times J}
\end{equation}
a) Show that, if $A$ and $B$ fulfil the KS condition then so too does $A\ast B$.

\ss
To every (finite) family $A=(k_i)_{i\in I}$ which fulfils the KS condition, we attach a transformation 
$\phi_A : \M_n\mapsto \M_n$, given by the formula 
\begin{equation}
\phi_A(M)=\sum_{i\in I} k_i M k_i^*	
\end{equation}
b) Show that $\phi_A$ is linear and preserves $\St_n$ (i.e. $\phi_A(\St_n)\subset \St_n$).\\
c) Show that if $A$ and $B$ fulfil the KS condition, one has 
\begin{equation}\label{composition_phi}
\phi_A\circ\phi_B=\phi_{A\ast B} 	
\end{equation}
Conclude that the $\phi_A$ form a semigroup of transformations (with unit).\\
d) (Example) Let $(E_{i,j})_{i,j\in\{1,2\}}$ be the set of the four matrix units in $\M_2$. Show that the following families fulfil KS condition
\begin{eqnarray}
A&=&(E_{11},\frac{1}{\sqrt(2)}E_{12},\frac{1}{\sqrt(2)}E_{22})\cr
B&=&(\frac{1}{\sqrt(2)}E_{11},\frac{1}{\sqrt(2)}E_{21},E_{22})	
\end{eqnarray}
Compute $A*B$.\\ 
3) (Description of the semigroup at the level of multisets). In order to pull-back the formula \mref{composition_phi} at the level of multisets, we remark that the order or the labelling of the elements $k_i$ is irrelevant; all that counts is their multiplicities.\\
a) (Example showing that the ``set'' structure is too weak). Let $(E_{i,j})_{i,j\in\{1,2\}}$ be the set of the four matrix units in $\M_2$ and set
\begin{eqnarray}
A&=&(\frac{1}{\sqrt(2)}E_{11},\frac{1}{\sqrt(2)}E_{12},\frac{1}{\sqrt(2)}E_{21},\frac{1}{\sqrt(2)}E_{22})
\end{eqnarray}
compute $A\ast A$ and check that it has (non-zero) repeated elements and thus corresponds to a multiset (see appendix).\\ 
b) Show that the multisets of $\M_n$ are exactly the elements $\sum_{M\in\M_n}\lambda_M [M]$ (here we note $[M]$ the image of $M\in \M_n$ in the algebra $\R[\M_n]$ in order to forbid matrix addition) of the algebra $\R[\M_n]$ such that 
\begin{equation}\label{multisets_in_algebra}
(\forall M\in\M_n)(\lambda_M\in\N)
\end{equation}
the set of elements fulfilling \mref{multisets_in_algebra} will be denoted $\N[\M_n]$.\\
b) To every finite family of matrices (in $\M_n$) $A=(M_i)_{i\in I}$ one may associate its sum (in $\R[\M_n]$) $S(A)=\sum_{i\in I}M_i$, check that it is an element of $\N[\M_n]$ and that every element of $\N[\M_n]$ is obtained so.\\
c) Show that $S(A\ast B)=S(A).S(B)\in\N[\M_n]\subset \R[\M_n]$ and deduce that $(\N[\M_n],.)$ is a monoid.\\
d) To every multiset of matrices (in $\M_n$) $A=\sum_{M\in\M_n}\al(M) [M]$, one associates 
$$
T(A)=\sum_{M\in\M_n}\al(M) [M^*M]
$$ 
show that, if $A=(M_i)_{i\in I}$ is a finite set of matrices, one has
\begin{equation}
\sum_{i\in I} M_i^*M_i=T(S(A))
\end{equation}
e) Show that, if $T(A)=I_n$, $T(A.B)=T(B)$.
\\
We denote $\N[\M_n]^{KS}$ the set of $A\in \N[\M_n]$ such that $T(A)=I_n$.\\
f) Check that the mapping $\phi_A$ defined previously depends only on $S(A)$ and denote, for $A\in \N[\M_n]^{KS}$ the mapping $\Phi_A$ deduced from this property.\\ 
g) Prove that the mapping $(\N[\M_n]^{KS}, .)\mapsto (End_{\C}(\C^{n\times n}),\circ)$ is a morphism of semigroups (preserving the units).

\ss 
4) (Invertible elements) To every $A=\sum_{M\in\M_n}\al(M) [M]\in \N[\M_n]^{KS}$ one associates $\ep(A)=\sum_{M\in\M_n}\al(M)\in \N$.\\ 
a) Prove that 
\begin{equation}
	\ep(A.B)=\ep(A)\ep(B)
\end{equation}
b) Prove that 
\begin{equation}
\ep(A)=1\Longrightarrow |supp(\al)|=1
\end{equation}
and, thus, in this condition, $A=[M]$ (a single matrix).\\
c) Deduce from (a) and (b) that the set of invertible elements of $\N[\M_n]^{KS}$ is exactly the unitary group $U(n)$. 
\end{exo}

\begin{exo}\label{quotient_monoid} Let $(M,\ast,1_M)$ be a monoid, $\equiv$ an equivalence relation on $M$ and $Cl : M\mapsto M/\equiv$ the class function (which, to every element of $M$ associates its class $Cl(x)$).\\
a) Suppose that there is an (internal) law $\perp$ on $M/\equiv$ such that $Cl$ is a morphism i.e.  one has 
\begin{equation}\label{quotient_monoid1}
(\forall x,y\in M) (Cl(x*y)=Cl(x)\perp Cl(y))	
\end{equation}
then prove that $\equiv$ is compatible with the right and left ``translations'' of the monoid this means that
\begin{equation}\label{congruence}
	(\forall\ x,y,t,s\in S)(x\equiv y\Longrightarrow [s*x*t\equiv s*y*t])
\end{equation}
b) Conversely, we suppose that $\equiv$ is an equivalence on $M$ satisfying condition \mref{congruence}, show that the result $Cl(x'*y')$ does not depend on the choice of $x'\in Cl(x);\ y'\in Cl(x)$ and therefore construct a law $\perp$ on $M/\equiv$ such that the class function is a morphism (i.e. \mref{quotient_monoid1}).\\
c) Show moreover that, in the preceding conditions, $\Big(M/\equiv,\perp,Cl(1_M)\Big)$ is a monoid.  
\end{exo}

\begin{exo}\label{tensor_rep} Let $\A$ be an AAU and $\Delta : \A\mapsto \A\otimes \A$ a comultiplication. We build (tensor) products of two representations by the formula \mref{tensor1}, more precisely by

\begin{equation}\label{tensor2}
can\circ (\rho_1\otimes \rho_2)\circ\Delta	
\end{equation}
where $can : End(V_1)\otimes End(V_2)\mapsto End(V_1\otimes V_2)$ is the canonical mapping.\\
a) Prove that, if $\Delta$ is a morphism of algebras, then the linear mapping 
\begin{equation}
\rho_1\sqtimes \rho_2 : \A\mapsto End(V_1\otimes V_2)
\end{equation}
defined by the composition \mref{tensor2} is a morphism of AAU (and hence a representation).\\
b) Prove that, if $\rho_1\sqtimes \rho_2$ is a representation for any pair $\rho_1,\rho_2$ of representations of $\A$, then $\Delta$ is a morphism of AAU (use $\rho_1=\rho_2$, the - regular - representation of $\A$ on itself by multiplications on the left).  
\end{exo}

\begin{exo}\label{tensor_assoc} We consider the canonical isomorphisms 
\begin{eqnarray}\label{trilinear_ident}
can_{1|23} : V_1\otimes (V_2\otimes V_3)\mapsto V_1\otimes V_2\otimes V_3\cr
can_{12|3} : (V_1\otimes V_2)\otimes V_3\mapsto V_1\otimes V_2\otimes V_3
\end{eqnarray}
\end{exo}
Show that, in order to have \mref{tensor_rep_assoc} for every triple $\rho_i,\ i=1,2,3$ of representations it is necessary and sufficient that 
\begin{equation}
can_{12|3}\circ(\Delta\otimes Id_\A)\circ \Delta=can_{1|23}\circ(Id_\A\otimes \Delta)\circ \Delta
\end{equation}
(for the necessary condition, consider again the left regular representations).

\begin{exo}\label{coalgebra_dual} 
Let $(\A,\Delta)$ be a coalgebra ($\Delta$ is an arbitrary - but fixed - linear mapping) and $(\A^*,\ast_\Delta)$ be its dual algebra. Explicitely, for $f,g\in \A^*$ and $x\in \A$ (for convenience, the law is written in infix denotation)
\begin{equation}
	\scal{f\ast_\Delta g}{x}=\scal{f\otimes g}{\Delta(x)}
\end{equation}
Prove the following equivalences
\begin{eqnarray}
	\Delta \text{ is co-associative }\Longleftrightarrow \ast_\Delta \text{ is associative }\\
\hspace{-2cm}	(\forall \ep\in \A^*)\Big(\ep \text{ is a unity for }(\A^*,\ast_\Delta)\Longleftrightarrow 
	\ep \text{ is a co-unity for }(\A,\Delta)\Big)
\end{eqnarray} 
\end{exo}

\begin{exo}\label{tensor_unit} The mappings $can_l,can_r$ are as in \mref{can_rl}. Prove that, in order that for any representation $\rho$ of $\A$, one has
\begin{equation}
	can_l\circ(\ep\otimes \rho)\circ\Delta=can_r\circ(\rho\otimes \ep)\circ\Delta
\end{equation}
it is necessary and sufficient that $\ep$ be a counit.
\end{exo}

\begin{exo}\label{convolution}
Let $(\c,\Delta)$ be a coalgebra and $(\A,\mu)$ ba an algebra on the same (commutative) field of scalars $k$. We define a multiplication (called convolution) on $Hom_k(\c,\A)$ by
\begin{equation}
	f*g=\mu\circ (f\otimes g)\circ \Delta
\end{equation}
so that, if $\Delta(x)=\sum_{(1)(2)}x_{(1)}x_{(2)}$, 
$$
f*g(x)=\sum_{(1)(2)}f(x_{(1)})g(x_{(2)}).
$$
1) If $\A$ is associative and $\c$ coassociative, show that the algebra $(Hom_k(\c,\A),*)$ is associative.\\
2) We suppose moreover that $\c$ admits a counit  
 $\ep : \c\mapsto k$ and $\A$ a unit $1_\A$ (identified with the linear mapping $k\mapsto \A$ given by $\lambda\ra \lambda 1_\A$).\\
Show that $1_\A\circ \ep$ (traditionnally denoted $1_\A\ep$) is the unit of the algebra $(Hom_k(\c,\A),*)$.\\
3) Let $(\B,\ast,1,\Delta,\ep)$ be a bialgebra. The convolution under consideration will be that constructed between the coalgebra $(\B,\Delta,\ep)$ and the algebra $(\B,\ast,1)$.\\ 
a) Let $S\in End(\B)$. Show that the following are equivalent\\ 
i) $S$ is an antipode for $\B$\\
ii) $S$ is the inverse of $Id_\B$ in $(End_k(\B),*)$.\\
b) Deduce from (b) that the antipode, if it exists, is unique.\\
c) Prove that the bialgebra $(\ncp{k}{A},\ast,\Delta_h,\ep_{aug})$ defined around equation \mref{letter_grouplike} admits no antipode (if the alphabet $A$ is not empty).\\ 
4) Let $(\c,\Delta,\ep)$ be a coalgebra coassociative with counit. We define $\Delta_2$ by $T_{2,3}\circ \Delta\otimes \Delta$ where $T_{2,3} : \c^{\otimes 4}\mapsto \c^{\otimes 4}$ is the flip between the 2nd and the 3rd component
\begin{equation}
T_{2,3}(x_1\otimes x_2\otimes x_3\otimes x_4)=x_1\otimes x_3\otimes x_2\otimes x_4
\end{equation}
a) Show that $(\c\otimes\c,\Delta_2,\ep\otimes\ep)$ (with $\ep\otimes\ep(x\otimes y)=\ep(x)\ep(y)$) is coassociative coalgebra with counit.\\
Let $(\H,\mu,1,\Delta,\ep,S)$ be a Hopf algebra. The convolution $*$ here will be that constructed between the coalgebra $(\H\otimes\H,\Delta_2,\ep\otimes\ep)$ and the algebra $(\H,\mu,1)$. We consider the two elements $\nu_i\in (Hom_k(\H\otimes\H,\H),*)$ defined by $\nu_1=S\circ\mu$ and $\nu_2(x\otimes y)=S(y)S(x)$.      
b) Show that the elements $\nu_i$ are the convolutional inverses of $\mu$. Deduce from this that $S : \H\mapsto \H$ is an antimorphism of algebras.
\end{exo}

\begin{exo}\label{locally_finite}
1) Let $(\B,\ast,1,\Delta,\ep)$ be a bialgebra, we denote by $\B^+$ the kernel of $\ep$.\\
a) Prove that $\B=\B^+\oplus k.1_\B$.\\
We denote $I^+$ the projection $\B\mapsto\B^+$ with respect to the preceding decomposition.\\
b) Prove that, for every $x\in\B^+$, one can write
\begin{equation}
	\Delta(x)=x\otimes 1+ 1\otimes x+ \sum_{(1)(2)}x_{(1)}\otimes x_{(2)} \text{ with } x_{(i)}\in \B^+
\end{equation}
2) Define for $x\in \B^+$, 
\begin{equation}
\Delta^+(x)=\Delta(x)-(x\otimes 1+ 1\otimes x)=\sum_{(1)(2)}x_{(1)}\otimes x_{(2)}
\end{equation}
a) Check that $(\B^+,\Delta^+)$ is a coassociative coalgebra.\\
Define
\begin{equation}
(\B^*)^+=\{f\in \B^*|f(1)=0\}	
\end{equation}
b) Prove that $(\B^*)^+$ is a subalgebra of $(\B,\ast_\Delta)$ and that its law is dual of $\Delta^+$.\\
c) Prove that the algebra $(\B^*,\ast_\Delta)$ is obtained from $\Big((\B^*)^+,\ast_{\Delta^+}\Big)$ by adjunction of the unity $\ep$.\\
3) The bialgebra is called locally finite if                                                                                                                  
\begin{equation}
	(\forall x\in \B)(\exists k\in \N^*)(\Delta^{+(k)}(x)=0).
\end{equation}
The projection $I^+$ being as above, show that, in case $\B$ is locally finite,
\begin{equation} 
(\forall x\in \B)(\exists N\in \N^*)(\forall k\geq N)((I^+)^k(x)=0) 
\end{equation}
and that 
\begin{equation}\label{antipode_by_series1}
	S=\sum_{n\in \N}(-I^+)^n
\end{equation}
is an antipode for $\B$.
\end{exo}
\begin{exo}
1) Let $G$ be a group and $\H=(\C[G],.,1_G,\Delta,\ep,S)$ be the Hopf algebra of $G$.\\
a) Show that $\{(g-1)\}_{g\in G-\{1\}}$ is a basis of $\H^+$ (defined as above) and that $\Delta^+(g-1)=(g-1)\otimes (g-1)$.\\
b) Show that, if $G\not=\{1\}$, $\H^+$ is not locally finite, but $\H$ admits an antipode.\\
2) Prove that, if the coproduct of $H$ is graded (i.e. there exists a decomposition $H=\oplus_{n\in \N}H_n$ with $\Delta(H_n)\subset \sum_{a+b=n} H_a\otimes H_b$) and $H_0=k.1_H$, then the comultiplication is locally finite.\\
3) Define the degree of a labelled diagram as its number of edges and $\LDIAG_n$ as the vector space generated by the diagrams of degree $n$ and check that we satisfy the conditions of exercise \mref{locally_finite} question 5. 
\end{exo}

\begin{exo}\label{struct_const_ex} 1) Show that, in order that a family $(\lambda_{i,j}^k)_{i,j,k\in I}$ be the family of structure constants of some algebra it is necessary and sufficient that 
\begin{equation}\label{struct_const_alg}
	(\forall (i,j)\in I^2)\Big((\lambda_{i,j}^k)_{k\in I}\textrm { is finitely supported}\Big)
\end{equation}
2) Similarly show that in order that a family $(\lambda_{i}^{j,k})_{i,j,k\in I}$ be the family of structure constants of some coalgebra it is necessary and sufficient that 
\begin{equation}\label{struct_const_coalg}
	(\forall i\in I)\Big((\lambda_{i}^{j,k})_{(j,k)\in I^2}\textrm { is finitely supported}\Big)
\end{equation}
3) Give examples of mappings $\lambda : I^3\mapsto k$ such that the corresponding families satisfy
\begin{itemize}
	\item[i)] \mref{struct_const_alg} and \mref{struct_const_coalg} 
	\item[ii)] \mref{struct_const_alg} and not \mref{struct_const_coalg} 
	\item[iii)] \mref{struct_const_coalg} and not \mref{struct_const_alg}  
	\item[iv)] none of \mref{struct_const_alg} and \mref{struct_const_coalg} 
\end{itemize}

4) Give further examples such as those in 3) i-iii but now defining associative (resp. coassociative) multiplications (resp. comultiplications).
\end{exo}

\section{APPENDIX}

\subsection{Function spaces}\label{function_spaces}

Throughout the text, we use the basic constructions of set theory and algebra (see \cite{B_Set,B_Alg_III}).\\
The set of mappings between two sets $X$ and $Y$ is denoted by $Y^X$. Thus if $k$ is a field 
\begin{equation}
k^X=\{f:X\longrightarrow k\}	
\end{equation}
the vector space of all functions defined on $X$ with values in $k$. For each function $f\in k^X$, we call the ``support of $f$'' the set of points $x\in X$ such that $f(x)$ is not zero\footnote{In integration theory, the support of a function is the closure of what we define as the (algebraic) support.}.  
\begin{equation}\label{support}
supp(f)=\{x\in X:f(x)\neq 0\}	
\end{equation}
the set of functions with finite support is a vector subspace of $k^X$ which is denoted by $k^{(X)}$. 

\ss
An interesting extension of this notion to other sets of coefficients is the combinatorial notion of (finite) multisets.\\
Recall that a multiset is a {\rm (set with repetitions)} \cite{K2}. For example, the first multisets with elements from $\{a,b\}$ are
\begin{eqnarray}\label{first_multisets}
\{\},\{a\},\{b\},\{a,a\},\{a,b\},\{b,b\},\{a,a,a\},\{a,a,b\},\{a,b,b\},\{b,b,b\},\cr
\{a,a,a,a\},\{a,a,a,b\},\{a,a,b,b\},\{a,b,b,b\},\{b,b,b,b\},\cdots .
\end{eqnarray}
A multiset with elements in $X$ is then described equivalently by a multiplicity function $\al : X\mapsto \N$ with finite support. The support of such a mapping is defined as in \mref{support} and the set of finite multiplicity functions will be denoted by $\N^{(X)}$ (see below the free commutative monoid \mref{fc_monoid}). For example, the multiplicity functions $\{a,b\}\mapsto \N$ corresponding to the multisets given in \mref{first_multisets} are, in the same order (we characterize $\al$ by the pair $(\al(a),\al(b))$)
\begin{eqnarray}\label{first_multisets}
(0,0),(1,0),(0,1),(2,0),(1,1),(0,2),(3,0),(2,1),(1,2),(0,3),\cr
(4,0),(3,1),(2,2),(1,3),(0,4),\cdots 
\end{eqnarray}

\subsection{Basic structures}\label{basic_structures}

\begin{definition} (Semigroup) A semigroup $(S,*)$ is a set $S$ endowed with a closed binary operation $*$ satisfying an associative law, this means that, for all 
$x,\ y,\ z\in S$ one has $x*(y*z)=(x*y)*z$. 
\end{definition}

{\tt http://en.wikipedia.org/wiki/Semigroup}

\begin{definition} (Monoid) A monoid $(M,\ast)$ is a semigroup which possesses a neutral element, i.e. an element $e\in M$ such that, for all $x\in M$:
\begin{eqnarray}
e\ast x=x\ast e=x.
\end{eqnarray}
Such an element, if it exists is unique. The neutral element is often denoted $1_M$.
\end{definition}

{\tt http://en.wikipedia.org/wiki/Monoid}

\begin{definition}\label{f_monoid} (Free Monoid) The free monoid of alphabet $X$ is the set of {\it strings} $x_1x_2\cdots x_n$ with letters $x_i\in X$ (comprising the empty string). This set is denoted $X^*$, its law is the concatenation and its neutral element is the empty string. 
\end{definition}

It is easily seen that this monoid is free in the following sense. For any ``set-theoretical'' mapping $\phi : X\mapsto M$, where $(M,\ast)$ is a monoid, $\phi$ can be extended to strings so that 
\begin{equation}\label{f_monoid_universal}
\begin{array}{c}\xymatrix{
\ar[r]^{\ \phi}X\ar[dr]_{can}& M\\
 &\ar[u]^{\bar\phi}X^*\\
}
\end{array}
\end{equation}

\begin{definition}\label{fc_monoid} (Free Commutative Monoid) The free commutative monoid of the alphabet $X$ is the set of {\it monomials} $X^\al$ ($\al\in \N^{(X)}$). This set is denoted by $\MON(X)$ and its law is the multiplication of monomials
\begin{equation}
X^\al X^\be=X^{\al+\be}	
\end{equation}
\end{definition}

It is easily seen that this monoid is free in the following sense. For any ``set-theoretical'' mapping $\phi : X\mapsto M$, where $(M,\ast)$ is a commutative monoid, $\phi$ can be extended to monomials so that 
\begin{equation}
\begin{array}{c}\xymatrix{
\ar[r]^{\ \phi}X\ar[dr]_{can}& M\\
 &\ar[u]^{\bar\phi} \MON(X)\\
}
\end{array}
\end{equation}

An interesting application of the free monoid is the explicit construction of a monoid defined ``by generators and relations''. 

Let $X$ be a set (of generators) and $R=(u_i,v_i)_{i\in I}$ a family of pairs of words, then one can construct explicitely the smallest (i.e. the intersection of) congruence $\equiv_R$ for which 
$$
(\forall i\in I) (u_i\equiv v_i). 
$$ 
Let say that two words $U,V\in X^*$ are ``related'' by $\equiv_R$ if there is a chain of replacements of the type 
$pu_is\ra pv_is$ or $pv_is\ra pu_is$ $p,s\in X^*$ leading from $U$ to $V$. Formally, there exists a chain 
\begin{equation}
	U=U_0,U_1,\cdots U_n=V
\end{equation}
such that for each $j<n$ $U_j=p_jAq_j\ ;\ U_{j+1}=p_jBq_j$ with $(A,B)=(u_i,v_i)$ or $(B,A)=(u_i,v_i)$ for some $i$ (depending on $j$. One can show that the constructed $\equiv_R$ is a congruence (see exercise \mref{congruence}) and we define
\begin{equation}
\langle X;R\rangle_{\bf Mon}	
\end{equation}
as the quotient $X^*/\equiv_R$. This monoid has the following property : if $\phi X^* \mapsto M$ is a morphism (of monoids) such that, for all $i\in I$ one has $\phi(u_i)=\phi(v_i)$, then $\phi$ factorises uniquely through 
$\langle X;R\rangle_{\bf Mon}$
 
\begin{equation}\label{gen_rel_universal}
\begin{array}{c}\xymatrix{
\ar[r]^{\ \phi}X\ar[dr]_{can}& M\\
 &\ar[u]^{\bar\phi} \langle X;R\rangle_{\bf Mon}\\
}
\end{array}
\end{equation}

\begin{definition} (Group) A group $(G,\ast)$ is a monoid such that for each $x\in G$ there exists $y$  such that
\begin{eqnarray}
x\ast y=y\ast x=e.
\end{eqnarray}
For fixed $x$ such an element is unique and is usually denoted by $x^{-1}$ and called the inverse of $x$.
\end{definition}

\begin{definition} (Algebra of a monoid) Let $k$ be a field (scalars, for example $k=\mathbb{R}$ or $\mathbb{C}$). The algebra $k[M]$ of a monoid $M$ (with coefficients in $k$) is the set of mappings $k^{(M)}$ endowed with the convolution product
\begin{equation}
	f\ast g(w)=\sum_{uv=w}f(u)g(v).
\end{equation}
The algebra $(k[M],\ast)$ is an AAU.
\end{definition}
Each $m\in M$ may be identified with its characteristic function (i.e. the Dirac function $\delta_m$ with value $1$ at $m$ and $0$ elsewhere). These functions form a basis of $k[M]$ and then, every $f\in k[M]$ can be written as a finite sum $f=\sum_w f(w)w$. Through this identification the unity of $M$ and $k[M]$ coincide.

The algebra of a monoid solves the following univeral problem.\\ 
Let $M$ be a monoid and $j:M\mapsto k[M]$ the embedding described above. For any AAU $\A$ and any morphism of monoids $\phi: M\mapsto \A$ ($(\A,.)$ is a monoid), one has an unique factorization 

\begin{eqnarray}\nonumber
\ \ \ \ \ \ \xymatrix{
M\ar[r]^{\phi}\ar[d]^{j}& \A\\
 k[M]\ar[ur]_{\bar{\phi}} &\\
}\end{eqnarray}
where $\bar{\phi}$ is a morphism of AAU.

Likewise, the enveloping algebra $\mathcal{U}_k(\mathfrak{G})$ of a Lie algebra $\mathfrak{G}$ with the canonical mapping $can: \mathcal{U}_k(\mathfrak{G})\mapsto \mathfrak{G}$ is the solution of a universal problem. The specifications are the following 
\begin{enumerate}
	\item $\mathcal{U}_k(\mathfrak{G})$ is an AAU
	\item $can$ is a morphism of Lie algebras (for this, $\mathcal{U}_k(\mathfrak{G})$ is endowed of the structure of Lie algebra given by the bracket $[X,Y]=XY-YX$).
\end{enumerate}
For any morphism of Lie algebras $\phi : \mathfrak{G}\mapsto \A$ (where $\A$ is endowed with the structure of a Lie algebra given by the bracket $[X,Y]=XY-YX$), one has a unique factorization   

\begin{eqnarray}
\xymatrix{
\mathfrak{G}\ar[r]^{\phi}\ar[d]^{can}& \A\\
 \mathcal{U}_k(\mathfrak{G})\ar[ur]_{\bar{\phi}} &\\
}\end{eqnarray}

where $\bar{\phi}$ is a morphism of AAU.

\end{document}